\documentclass{PoS}

\usepackage[centertags]{amsmath}
\usepackage{amsfonts}
\usepackage{amssymb}
\usepackage{amsthm}
\usepackage{amsmath}
\usepackage{graphicx,color}
\usepackage{ulem}
\usepackage{feynmf}

\newcommand{\beq}{\begin{equation}}
\newcommand{\eeq}[1]{\label{#1}\end{equation}}
\newcommand{\bea}{\begin{eqnarray}}
\newcommand{\eea}[1]{\label{#1}\end{eqnarray}}

\newcommand{\ba}[1]{\begin{align}\label{#1}}
\newcommand{\ea}{\end{align}}

\def\bs{\begin{split}}
\def\es{\end{split}}
\def\ba{\begin{array}}
\def\ea{\end{array}}
\def\bec{\begin{center}}
\def\ec{\end{center}}
\def\ba{\begin{align}}
\def\ena{\end{align}}

\def\de{\partial}

\def\gz0{\gamma^{0}}

\def\nn{\nonumber}

\def\a{\alpha}
\def\b{\beta}

\def\e{\epsilon}

\def\l{\lambda}
\def\L{\Lambda}
\def\m{\mu}
\def\n{\nu}

\def\p{\pi}

\def\r{\rho}

\def\s{\sigma}

\def\vf{\varphi}

\def\12{\frac{1}{2}}

\def\dag{\dagger}

\title{Higher Spin Theory - Part I}

\ShortTitle{Higher Spin Theory - Part I}

\author{\speaker{Rakibur Rahman}\\
        Physique Th\'eorique et Math\'ematique \& International Solvay Institues\\
        Universit\'e Libre de Bruxelles, Campus Plaine C.P. 231, B-1050 Bruxelles, Belgium\\
        E-mail: \email{rakibur.rahman@ulb.ac.be}}

\abstract{These notes comprise a part of the introductory lectures on Higher Spin Theory presented in the Eighth Modave Summer School in Mathematical Physics.
We construct free higher-spin theories and turn on interactions to find that
inconsistencies show up in general. Interacting massless fields in flat space are in tension with gauge invariance and this leads to various no-go theorems.
While massive fields exhibit superluminal propagation, appropriate non-minimal terms may cure such pathologies as they do in String Theory$-$a fact that we
demonstrate. Given that any interacting massive higher-spin particle is described by an effective field theory, we compute a model independent upper bound
on the ultraviolet cutoff in the case of electromagnetic coupling in flat space and discuss its implications. Finally, we consider various possibilities
of evading the no-go theorems for massless fields, among which Vasiliev's higher-spin gauge theory is one. We employ the BRST-antifield method for a simple
but non-trivial gauge system in flat space to find a non-abelian cubic coupling and to explore its higher-order consistency.}

\FullConference{Eighth Modave Summer School in Mathematical Physics,\\
		26th august - 1st september 2012\\
		Modave, Belgium }

\begin{document}

\section{Motivations \& Outline}

In Quantum Field Theory fundamental particles carry irreducible unitary representations of the Poincar\'{e} group,
and therefore can have arbitrary (integer or half-integer) values of the spin, at least in principle. The motivations
for studying higher-spin (HS) particles are manifold.
\begin{enumerate}
\item
While free HS fields are fine, severe problems show up as soon as interactions are turned on. For massless particles,
there exist powerful no-go theorems~\cite{w64,gpv,ad,ww,p} that forbid them from interacting in flat space with electromagnetism
(EM) or gravity when their spin $s$ exceeds certain value. One would like to evade these theorems in order write down interacting
theories for massless HS fields.
\item
Massive HS particles do exist in the form of hadronic resonances, e.g., $\pi_2$(1670), $\rho_3$(1690), $a_4$(2040) etc.
These particles are composite, and their interactions are described by complicated form factors. However, when the exchanged
momenta are small compared to their masses, one should be able to describe their dynamics by consistent local actions.
\item
Massive HS excitations show up in (open) string spectra. In fact, they play a crucial role in that some of the spectacular features
of String Theory, e.g., planar duality, modular invariance and open-closed duality, rest heavily on their presence.
\item
It is conjectured that String Theory describes a broken phase of a HS gauge theory~\cite{Gross,Amati,Vasiliev0,Sagnotti11}.
The underlying HS theory may provide a better understanding of what String Theory is.
\item
Vasiliev's HS gauge theory in $\text{AdS}_4$~\cite{Vasiliev0,Vasiliev1} is conjectured to be holographically dual to
$O(N)$ Vector Models~\cite{KP}, and the first non-trivial checks of the duality appear in~\cite{GiombiYin}.
This duality may be regarded as the simplest non-trivial example of AdS/CFT correspondence, and may help us understand
some aspects of gauge/gravity dualities in general.
\end{enumerate}

In what follows we will present an introduction to HS fields and their interactions. In Section~\ref{sec:two}, we start
with construction of free HS theories for massive and massless fields. In  Section~\ref{sec:three}, we discuss the
difficulties that one faces when interactions are turned on. In particular, Section~\ref{sec:no-go} presents the various
no-go theorems~\cite{w64,gpv,ad,ww,p} for interacting HS massless fields in flat space, all of which essentially derive from
the fact that interactions are in tension with gauge invariance, while Section~\ref{sec:vz} shows that massive HS fields exhibit
superluminal propagation in external backgrounds$-$the notorious Velo-Zwanziger problem~\cite{vz}. Appropriate non-minimal
terms may cure the Velo-Zwanziger acausality, and in Section~\ref{sec:open} we show how String Theory provides remedy to
this pathology. In Section~\ref{sec:five}, we consider the fact that interacting massive HS particles are described
by effective field theories, compute a model independent upper bound on the ultraviolet cutoff for their electromagnetic
coupling in flat space, and discuss the implications of this result. Finally, in Section~\ref{sec:six} we study interactions
of HS gauge fields by considering the different possibilities of evading the aforementioned no-go theorems, e.g., adding a
mass term, working in $D=3$ or in AdS space, or considering higher-derivative interactions. We work out
a simple example of a $1-\tfrac{3}{2}-\tfrac{3}{2}$ non-abelain cubic coupling in flat space by using the BRST-antifield
formalism and explore its higher-order consistency.

\section{Construction of Free Higher Spin Theories}\label{sec:two}

The task of constructing a theory of HS fields dates back to 1936, when Dirac tried to generalize
his celebrated spin-$\tfrac{1}{2}$ equation~\cite{dirac}. A systematic study of HS particles, though, was
initiated by Fierz and Pauli in 1939~\cite{pf}. Their approach was field theoretic that focused on the physical
requirements of Lorentz invariance and positivity of energy. The works of Wigner on unitary representations of
the Poincar\'{e} group~\cite{wig1} and of Bargmann and Wigner on relativistic wave equations~\cite{wig2},
however, made it clear that one could replace positivity of energy by the requirement that a one-particle
state carries an irreducible unitary representation of the Poincar\'{e} group.

\subsection{Massive Fields}

The two Casimir invariants of the Poincar\'{e} group are \beq C_1=P^\mu P_\mu,\qquad C_2=W^\mu W_\mu,\eeq{Casimirs}
where $W^\mu\equiv-\tfrac{1}{2}\varepsilon^{\mu\nu\rho\sigma}J_{\nu\rho}P_\sigma$ is the Pauli-Lubanski pseudovector.
They define respectively mass and spin$-$the two basic quantum numbers$-$a field may possess. Let us consider the
first Wigner class, which corresponds to a physical massive particle of mass $m$ and spin $s$. In this case, we
have $C_1=-m^2$ and $C_2=m^2s(s+1)$. The transformation properties under the Lorentz group is uniquely determined
for a bosonic field by the usual choice of the representation $D\left(\tfrac{s}{2},\tfrac{s}{2}\right)$. The
field is then a totally symmetric rank-$s$ tensor, $\phi_{\mu_1 ... \mu_s}$, which is traceless:
\beq \eta^{\mu_1\mu_2}\phi_{\mu_1\mu_2 ... \mu_s}=0.\eeq{traceless}
The Casimir $C_1$ then demands that the Klein-Gordon equation be satisfied:
\beq (\Box-m^2)\phi_{\mu_1 ... \mu_s}=0.\eeq{h1}
The representation $D\left(\tfrac{s}{2},\tfrac{s}{2}\right)$ is however reducible under the rotation
subgroup, because it contains all spin values from $s$ down to $0$. The Casimir $C_2$ requires that all lower spin
values be eliminated; this is done by imposing the divergence/transversality condition:
\beq \partial^{\mu_1}\phi_{\mu_1 ... \mu_s}=0.\eeq{h2}
The latter is a necessary condition in order for the total energy to be positive definite~\cite{pf}.

How many degrees of freedom (DoF) does this field propagate in $D$ space-time dimensions? Note that the number
of independent components of symmetric rank-$s$ tensor is given by
\beq C(D+s-1,\,s)=\frac{(D-1+s)!}{s!(D-1)!}\,.\eeq{binomail} The tracelessness condition is a
symmetric rank-$(s-2)$ tensor that removes $C(D-3+s,\,s-2)$ of these components. Similarly, the
divergence constraint~(\ref{h2}) should eliminate $C(D-2+s,\,s-1)$ many, but its trace part
has already been incorporated in the tracelessness of the field itself, so that the actual number
is less by $C(D-4+s,\,s-3)$. Then the total number of propagating DoFs are
\bea \mathfrak{D}_{b,\,m\neq0}&=&C(D-1+s,\,s)-C(D-3+s,\,s-2)-C(D-2+s,\,s-1)+C(D-4+s,\,s-3)\nonumber\\
&=&C(D-4+s,\,s)+2\,C(D-4+s,\,s-1).\eea{dofBm} In particular when $D=4$, this number reduces to $2s+1$
as expected.

A fermionic field of spin $s=n+\tfrac{1}{2}$, on the other hand, is represented by a totally symmetric tensor-spinor
of rank $n$, $\psi_{\mu_1 ... \mu_n}$, which is $\gamma$-traceless:
\beq \gamma^{\mu_1}\psi_{\mu_1 ... \mu_n}=0.\eeq{gamma-traceless}
It transforms as the $D\left(\tfrac{n+1}{2},\tfrac{n}{2}\right) \bigoplus D\left(\tfrac{n}{2},\tfrac{n+1}{2}\right)$
representation of the Lorentz group , and satisfies
\bea (\not{\!\partial\!}-m)\psi_{\mu_1 ... \mu_n}&=&0,\label{h3}\\\partial^{\mu_1}\psi_{\mu_1 ... \mu_n}&=&0.\eea{h4}

The counting of DoFs is analogous to that for bosons. In $D$ dimensions, a symmetric rank-$n$ tensor-spinor
has $C(D+n-1,\,n)\times 2^{[D]/2}$ independent components, where $[D]\equiv D+\tfrac{1}{2}\left[(-1)^D-1\right]$.
The $\gamma$-trace and divergence constraints, being symmetric rank-$(n-1)$ tensor-spinors, each eliminates
$C(D+n-2,\,n-1)\times 2^{[D]/2}$ of them. But there is an over-counting of $C(D+n-3,\,n-2)\times 2^{[D]/2}$
constraints, since only the traceless part of the divergence condition~(\ref{h4}) should be counted.
The number of dynamical DoFs$-$fields and conjugate momenta$-$is therefore
\bea \mathfrak{D}_{f,\,m\neq0}&=&\left[C(D+n-1,\,n)-2\,C(D+n-2,\,n-1)+C(D+n-3,\,n-2)\right]\times 2^{[D]/2}\nonumber\\
&=&C(D-3+n,\,n)\times 2^{[D]/2}.\eea{dofFm} In $D=4$ in particular, this is $2n+2=2s+1$ as expected.

Fierz and Pauli noted in~\cite{pf} that attempts to introduce electromagnetic interaction at the level of the
equations of motion (EoM) and constraints lead to algebraic inconsistencies for spin greater than 1. If one
modifies the Eqs.~(\ref{traceless})-(\ref{h2}) and~(\ref{gamma-traceless})-(\ref{h4}) by directly replacing
ordinary derivatives with covariant ones, they are no longer mutually compatible\footnote{We will work out an example
of this inconsistency in Section~\ref{sec:three}.}. To avoid such difficulties it was suggested that all
equations involving derivatives be obtainable from a Lagrangian. This is possible only at the cost of introducing
lower-spin auxiliary fields, which must vanish when interactions are absent.

Fronsdal~\cite{frons} and Chang~\cite{chang} spelled out a procedure for introducing these auxiliary fields to
construct HS Lagrangians. Singh and Hagen~\cite{sh} achieved the feat in 1974 by writing down an explicit form of
the Lagrangian for a free massive field of arbitrary spin. The Singh-Hagen Lagrangian for an integer spin $s$
is written in terms of a set of symmetric traceless tensor fields of rank $s,s-2,s-3,...,0$. For a
half-integer spin $s=n+\tfrac{1}{2}$, the Lagrangian incorporates symmetric $\gamma$-traceless tensor-spinors:
one of rank $n$, another of rank $n-1$, and doublets of rank $n-2,n-3,...,0$. When the Eqs.~(\ref{traceless})-(\ref{h2})
and~(\ref{gamma-traceless})-(\ref{h4}) are satisfied, all the lower spin fields are forced to vanish.

To understand the salient features of the Singh-Hagen construction, we consider the massive spin-2 field,
which is described by a symmetric traceless rank-2 tensor $\phi_{\mu\nu}$. Let us try to incorporate the
Klein-Gordon equation, $(\Box-m^2)\phi_{\mu\nu}=0$, and the transversality condition, $\partial^\mu\phi_{\mu\nu}=0$, into
a Lagrangian as \beq \mathcal L=-\tfrac{1}{2}(\partial_\mu\phi_{\nu\rho})^2 -\tfrac{1}{2}m^2
\phi_{\mu\nu}^2 +\tfrac{1}{2}\alpha(\partial^\mu\phi_{\mu\nu})^2, \eeq{h5} where the constant $\alpha$ is to be
determined. While varying the above action one must keep in mind that $\phi_{\mu\nu}$ is symmetric traceless. One
therefore obtains the following EoMs: \beq (\Box-m^2)\phi_{\mu\nu}-\tfrac{1}{2}\alpha
\left[\partial^\rho\partial_{(\mu}\phi_{\nu)\rho}-\tfrac{2}{D}\,\eta_{\mu\nu}\partial^\rho\partial^\sigma
\phi_{\rho\sigma}\right]=0, \eeq{h6} where $D$ is the space-time dimensionality. The divergence of Eq.~(\ref{h6})
gives \beq [(\alpha-2)\Box+2m^2]\partial^\mu\phi_{\mu\nu}+\alpha\left(1-\tfrac{2}{D}\right) \partial_\nu
\partial^\rho\partial^\sigma\phi_{\rho\sigma}=0.\eeq{h7} The transversality condition can be recovered
by setting $\alpha=2$ as well as requiring $\partial^\mu\partial^\nu\phi_{\mu\nu}=0$. The latter condition,
however, does not follow from the EoMs. This problem can be taken care of by introducing an auxiliary spin-0
field $\phi$, so that the condition $\partial^\mu\partial^\nu\phi_{\mu\nu}=0$ follows from the Lagrangian.
Let us add the following terms to the Lagrangian~(\ref{h5}): \beq \mathcal L_{\text{add}} = \beta_1(\partial_\mu\phi)^2
+\beta_2\phi^2+\phi\,\partial^\mu\partial^\nu\phi_{\mu\nu},\eeq{h8} where $\beta_{1,2}$ are constants. The double
divergence of the $\phi_{\mu\nu}$ EoMs now gives \beq [(2-D)\Box-Dm^2]\partial^\mu\partial^\nu\phi_{\mu\nu}+(D-1)\Box\Box\phi=0.
\eeq{h9} The EoM for the auxiliary scalar $\phi$, on the other hand, reduces to \beq \partial^\mu\partial^\nu
\phi_{\mu\nu}-2(\beta_1\Box-\beta_2)\phi=0.\eeq{h10} Eqs.~(\ref{h9})-(\ref{h10}) comprise a linear homogeneous system
in the variables $\partial^\mu\partial^\nu\phi_{\mu\nu}$ and $\phi$. The condition $\partial^\mu
\partial^\nu\phi_{\mu\nu}=0$ and the vanishing of the auxiliary field must follow if the associated determinant
does not vanish. The latter is given by \beq \Delta=[2(D-2)\beta_1-(D-1)]\Box\Box+2[Dm^2\beta_1-(D-2)\beta_2]
\Box-2Dm^2\beta_2.\eeq{h11} Note that $\Delta$ becomes algebraic, not containing $\Box$, proportional to $m^2$
and hence non-zero if
\beq \beta_1=\tfrac{(D-1)}{2(D-2)}, \qquad\quad \beta_2=\tfrac{m^2D(D-1)}{2(D-2)^2}, \qquad\quad D>2.\eeq{h12}
Thus one constructs the following Lagrangian for a massive spin-2 field \beq \mathcal L=-\tfrac{1}{2}(\partial_\mu
\phi_{\nu\rho})^2\,+\,(\partial^\mu\phi_{\mu\nu})^2\,+\,\tfrac{(D-1)}{2(D-2)}(\partial_\mu\phi)^2
\,+\,\phi\,\partial^\mu\partial^\nu\phi_{\mu\nu}-\tfrac{1}{2}m^2\left[\phi_{\mu\nu}^2-\tfrac{D(D-1)}
{(D-2)^2}\,\phi^2\right].\eeq{h13} One can now check that this Lagrangian yields $\partial^\mu\partial^\nu\phi_{\mu\nu}=0$
and $\phi=0$, so that the Klein-Gordon equation and the transversality condition follow from it.

One can carry out this procedure for higher spins to find that the following pattern emerges: For an integer spin $s$,
one must successively obtain $\partial^{\mu_1}...\partial^{\mu_k}\phi_{\mu_1...\mu_s}=0$, for $k=2,3,...,s$. At each value
of $k$ one needs to introduce an auxiliary symmetric traceless rank-$(s-k)$ tensor field. Similarly for a
half-integer spin $s=n+\tfrac{1}{2}$, the transversality condition, $\partial^{\mu_1}\psi_{\mu_1 ... \mu_n}=0$, can be
obtained by introducing a symmetric $\gamma$-traceless tensor-spinor of rank $n-1$, $\psi_{\mu_1 ... \mu_{n-1}}$,
provided that one also satisfies the conditions: $\partial^{\mu_1}\partial^{\mu_2}\psi_{\mu_1 ... \mu_n}=0$ and
$\partial^{\mu_1}\psi_{\mu_1 ... \mu_{n-1}}=0$. These can be achieved by successively obtaining
$\partial^{\mu_1}...\partial^{\mu_k}\psi_{\mu_1...\mu_n}=0$ and $\partial^{\mu_1}...\partial^{\mu_k}
\psi_{\mu_1...\mu_{n-1}}=0$, for $k=2,3,...,n$. Then for each $k$ one must introduce two symmetric
$\gamma$-traceless rank-($n-k$) tensor-spinors. The explicit form of the Lagrangian for an arbitrary spin is
rather complicated and is given in Ref.~\cite{sh}.

\subsection{Massless Fields}

In 1978, Fronsdal and Fang~\cite{ff} considered the massless limit of the Singh-Hagen Lagrangian
to find considerable simplification of the theory. For the bosonic case, all the auxiliary fields
decouple in this limit, except for the one with the highest rank $s-2$. Furthermore, the two symmetric
traceless rank-$s$ and rank-$(s-2)$ tensors can be combined into a single symmetric tensor, $\Phi_{\mu_1...\mu_s}$,
which is double traceless: $\eta^{\mu_1\mu_2}\eta^{\mu_3\mu_4}\Phi_{\mu_1...\mu_s}=0$. The Lagrangian
reduces to~\cite{ff} \bea \mathcal L&=&-\tfrac{1}{2}(\partial_\rho
\Phi_{\mu_1...\mu_s})^2 +\tfrac{1}{2}s(\partial\cdot\Phi_{\mu_2...\mu_s})^2+\tfrac{1}{2}s(s-1)\left(
\partial\cdot\partial\cdot\Phi_{\mu_3...\mu_s}\right)\Phi'\,^{\mu_3...\mu_s}\nonumber\\&&+
\tfrac{1}{4}s(s-1)(\partial_\rho\Phi'_{\mu_3...\mu_s})^2+\tfrac{1}{8}s(s-1)(s-2)(\partial\cdot
\Phi'_{\mu_4...\mu_s})^2,\eea{b00} where prime denotes trace: $\Phi'_{\mu_3...\mu_s}\equiv\eta^{\mu_1\mu_2}\Phi_{\mu_1...\mu_s}$,
and $\partial\cdot$ denotes divergence: $\partial\cdot\Phi_{\mu_2...\mu_s}\equiv\partial^{\mu_1}\Phi_{\mu_1...\mu_s}$.
The Lagrangian~(\ref{b00}) has acquired a gauge invariance with a symmetric traceless gauge parameter
$\lambda_{\mu_1...\mu_{s-1}}$: \beq \delta\Phi_{\mu_1...\mu_s}=\partial_{(\mu_1}
\lambda_{\mu_2...\mu_s)},\qquad\quad \lambda'_{\mu_3...\mu_{s-1}}=0.\eeq{b01}

Let us illustrate the Fronsdal-Fang formulation~\cite{ff} by considering again the spin-2 case. In the massless limit,
the Singh-Hagen spin-2 Lagrangian~(\ref{h13}) reduces to \beq \mathcal L=-\tfrac{1}{2}(\partial_\mu\phi_{\nu\rho})^2\,
+\,(\partial\cdot\phi_\mu)^2\,+\,\tfrac{(D-1)}{2(D-2)}(\partial_\mu\phi)^2\,+\,\phi\,\partial\cdot
\partial\cdot\phi.\eeq{h135} We combine $\phi_{\mu\nu}$ and $\phi$ into a single field $h_{\mu\nu}$:
\beq h_{\mu\nu}\equiv\phi_{\mu\nu}+\tfrac{1}{D-2}\,\eta_{\mu\nu}\phi.\eeq{h14} The tracelessness
of $\phi_{\mu\nu}$ then gives \beq h'\equiv h^\mu_{~\mu}=\tfrac{D}{D-2}\,\phi.\eeq{h140} This reduces the Lagrangian~(\ref{h135})
to \beq \mathcal{L}=-\tfrac{1}{2}(\partial_\mu h_{\nu\rho})^2\,+\,(\partial\cdot h_\mu)^2\,+\,\tfrac{1}{2}(\partial_\mu
h')^2\,+\,h'\,\partial\cdot\partial\cdot h.\eeq{h15} Here the new field
$h_{\mu\nu}$ is symmetric but not traceless. The Lagrangian~(\ref{h15}) is nothing but the linearized
Einstein-Hilbert action with $h_{\mu\nu}$ identified as the metric perturbation around Minkowski background.
This describes a massless spin-2 particle, and the corresponding gauge symmetry is just the infinitesimal
version of coordinate transformations: \beq \delta h_{\mu\nu}=\partial_{(\mu}\lambda_{\nu)}.\eeq{h16}

Gauge invariance and the trace conditions on the field and the gauge parameter play a crucial role
in the Fronsdal-Fang construction in that they ensure absence of ghosts, and give the correct number
of propagating DoFs. To do the DoF count, we write down the EoMs that follow from the Lagrangian~(\ref{b00}):
\beq \mathcal F_{\mu_1...\mu_s}-\tfrac{1}{2}\eta_{(\mu_1\mu_2}\mathcal F'_{\mu_3...\mu_s)}=0,\eeq{abba0}
where $\mathcal F_{\mu_1...\mu_s}$ is the Fronsdal tensor defined as
\beq \mathcal F_{\mu_1...\mu_s}\equiv\Box\Phi_{\mu_1...\mu_s}-\partial_{(\mu_1}\partial\cdot\Phi_{\mu_2...\mu_s)}
+\partial_{(\mu_1}\partial_{\mu_2}\Phi'_{\mu_3...\mu_s)}.\eeq{abba1} Another equivalent form of the EoMs is the
so-called Fronsdal form: \beq \mathcal F_{\mu_1...\mu_s}=0.\eeq{abba2} This contains precisely the correct number of
conditions to determine the components of the symmetric double-traceless field $\Phi_{\mu_1...\mu_s}$
because $\mathcal F''_{\mu_5...\mu_s}$ is vanishing. A symmetric rank-$s$ tensor with vanishing double trace has
$C(D-1+s,\,s)-C(D-5+s,\,s-4)$ independent components in $D$ dimensions.
Now that Eq.~(\ref{abba1}) enjoys the gauge invariance~(\ref{b01}), the symmetric traceless rank-$(s-1)$ gauge
parameter enables us to remove $C(D-2+s,\,s-1)-C(D-4+s,\,s-3)$ components by imposing an appropriate covariant
gauge condition, e.g.,\footnote{This is the arbitrary-spin generalization of the Lorenz gauge for $s=1$ and the
de Donder gauge for $s=2$.} \beq \mathcal G_{\mu_2...\mu_s}\equiv\partial\cdot\Phi_{\mu_2...\mu_s}-\tfrac{1}{2}
\partial_{(\mu_2}\Phi'_{\mu_3...\mu_s)}=0,\eeq{abba3} where $\mathcal G_{\mu_2...\mu_s}$ is traceless because
$\Phi''_{\mu_5...\mu_s}=0$. This reduces Eq.~(\ref{abba2}) to \beq \Box\Phi_{\mu_1...\mu_s}=0.\eeq{abba4}
Thus $\Phi_{\mu_1...\mu_s}$ indeed describes a massless field. Eq.~(\ref{abba3}) however does not
completely fix the gauge since gauge parameters satisfying $\Box\lambda_{\mu_1...\mu_{s-1}}=0$ are still allowed.
Therefore, we can gauge away another set of $C(D-2+s,\,s-1)-C(D-4+s,\,s-3)$  components. Thus the total number of
propagating DoFs turns out to be~\cite{deW}:
\beq \mathfrak{D}_{b,\,m=0}=C(D-5+s,\,s)+2\,C(D-5+s,\,s-1).\eeq{dofB0}
In particular, this leaves us exactly with 2 DoF for all $s$ in $D=4$.

For the fermionic case too we can take the massless limit of the Singh-Hagen Lagrangian. For spin
$s=n+\tfrac{1}{2}$, one is left in this case only with three symmetric $\gamma$-traceless tensor-spinors
of rank $n,n-1$ and $n-2$ respectively; the other auxiliary fields decouple completely. These three can be combined into
a single symmetric rank-$n$ tensor-spinor, $\Psi_{\mu_1...\mu_n}$, which is triple $\gamma$-traceless:
$\gamma^{\mu_1}\gamma^{\mu_3}\gamma^{\mu_3}\Psi_{\mu_1...\mu_n}=0$. This leads to the following Lagrangian~\cite{ff}:
\bea i\mathcal L &=& \bar{\Psi}_{\mu_1...\mu_n}\not{\!\partial\!}\;\Psi^{\mu_1...\mu_n}+n\bar{\displaystyle\not{\!\!\Psi}}
_{\mu_2...\mu_n}\not{\!\partial\!}\,\displaystyle\not{\!\!\Psi}^{\mu_2...\mu_n}
-\tfrac{1}{4}n(n-1)\bar{\Psi}'_{\mu_3...\mu_n}\not{\!\partial\!}\;\Psi'^{\,\mu_3...\mu_n}\nonumber\\
&&-n\left[\bar{\displaystyle\not{\!\!\Psi}}_{\mu_2...\mu_n}\partial\cdot\Psi^{\mu_2...\mu_n}
-\text{h.c.}\right]+\tfrac{1}{2}n(n-1)\left[\bar{\Psi}'_{\mu_3...\mu_n}
\partial\cdot\displaystyle\not{\!\!\Psi}^{\,\mu_3...\mu_n}-\text{h.c.}\right].\eea{f00}
The resulting EoMs read
\beq \mathcal S_{\mu_1...\mu_n}-\tfrac{1}{2}\gamma_{(\mu_1}\displaystyle\not{\!\!\mathcal S}_{\mu_2...\mu_n)}
-\tfrac{1}{2}\eta_{(\mu_1\mu_2}\mathcal S^\prime_{\mu_3...\mu_n)}=0,\eeq{noname44} where the fermionic
Fronsdal tensor $\mathcal S_{\mu_1...\mu_n}$ is given by
\beq \mathcal S_{\mu_1...\mu_n}\equiv i\left[\displaystyle{\not{\!\partial\,}}\Psi_{\mu_1...\mu_n}
-\partial_{(\mu_1}\displaystyle{\not{\!\!\Psi}}_{\mu_2...\mu_n)}\right].\eeq{noname45}
One can also rewrite the EoMs in the Fronsdal form: \beq \mathcal S_{\mu_1...\mu_n}=0.\eeq{noname46}
This equation contains the appropriate number of independent components because $\mathcal S_{\mu_1...\mu_n}$
is triple $\gamma$-traceless if $\Psi_{\mu_1...\mu_n}$ is.
The Lagrangian~(\ref{f00}) and the EoMs~(\ref{noname44})-(\ref{noname46}) enjoy a gauge symmetry with a
symmetric $\gamma$-traceless tensor-spinor parameter $\varepsilon_{\mu_1...\mu_{n-1}}$:
\beq \delta\Psi_{\mu_1...\mu_n}=\partial_{(\mu_1}\varepsilon_{\mu_2...\mu_n)},\qquad\quad\gamma^{\mu_1}
\varepsilon_{\mu_1...\mu_{n-1}}=0.\eeq{f01}

To get the DoF count let us note that a symmetric triple $\gamma$-traceless rank-$n$ tensor-spinor
has $\left[C(D+n-1,\,n)-C(D+n-4,\,n-3)\right]\times 2^{[D]/2}$ components in $D$ dimensions.
However, the $\gamma$-traceless part of $\Psi_{0\mu_2...\mu_n}$ is actually non-dynamical as
Eq.~(\ref{noname46}) shows. Therefore, we have
$\left[C(D+n-2,\,n-1)-C(D+n-3,\,n-2)\right]\times 2^{[D]/2}$ constraints. Now, the symmetric $\gamma$-traceless
gauge parameter $\varepsilon_{\mu_1...\mu_{n-1}}$ enables us to choose, for example, the gauge\footnote{
Here the gauge condition does not involve derivatives, so that it cannot convert constraints into evolution
equations. This is in contrast with the bosonic case where the gauge condition~(\ref{abba3}) renders the
non-dynamical components$-$the traceless part of $\Phi_{0\mu_2...\mu_s}$$-$dynamical as seen from Eq.~(\ref{abba4}).}
\beq \mathcal G_{\mu_2...\mu_n}\equiv\displaystyle{\not{\!\!\Psi}}_{\mu_2...\mu_n}-\tfrac{1}{D-4+2n}\,
\gamma_{(\mu_2}\Psi'_{\mu_3...\mu_n)}=0,\eeq{abba7} where $\mathcal G_{\mu_2...\mu_n}$ is $\gamma$-traceless
because $\displaystyle{\not{\!\!\Psi}}_{\mu_4...\mu_n}^\prime=0$. This allows further gauge transformations
for which the gauge parameter satisfies
\beq \displaystyle{\not{\!\partial\,}}\varepsilon_{\mu_2...\mu_n}-\tfrac{2}{D-4+2n}\,
\gamma_{(\mu_2}\partial\cdot\varepsilon_{\mu_3...\mu_n)}=0.\eeq{abba8} Note that the left-hand side of
Eq.~(\ref{abba8}) is $\gamma$-traceless because $\varepsilon_{\mu_2...\mu_n}$ is. The gauge
condition~(\ref{abba7}) and the residual gauge fixing eliminate
$\left[C(D+n-2,\,n-1)-C(D+n-3,\,n-2)\right]\times 2^{[D]/2}$ components each.
In this way, one finds that the number of dynamical DoFs$-$fields and conjugate momenta$-$is~\cite{deW}:
\beq \mathfrak{D}_{f,\,m=0}=C(D-4+n,\,n)\times 2^{[D]/2}.\eeq{dofF} In $D=4$ in particular, this is 4 as expected.
We parenthetically remark that the gauge fixing~(\ref{abba7}) does not reduce the EoMs~(\ref{noname46}) to the Dirac
equation. However, one can use part of the residual gauge invariance, which satisfies the condition~(\ref{abba8}),
to set $\Psi'_{\mu_3...\mu_n}=0$. Then the field equations become $\displaystyle{\not{\!\partial\,}}\Psi_{\mu_1...\mu_n}=0$,
which indeed describe a massless fermion.

Apart from their gauge theoretic and geometric aspects\footnote{One can construct HS analogs of the spin-2 Christoffel
connection~\cite{deW}$-$a hierarchy of generalized Christoffel symbols linear in derivatives of the field with simple gauge
transformation properties. The wave equations are also very simple in terms of these objects. The trace conditions on the HS
field and the gauge parameter can also be easily understood in this approach. See also Ref.~\cite{Dario} for geometric
formulations of HS gauge fields.}, massless HS theories are interesting since
they give rise to the massive theories through proper Kaluza-Klein (KK) reductions~\cite{ady}. This construction$-$much simpler
than the original Singh-Hagen one$-$justifies the choice of auxiliary fields in the latter. For the bosonic case, for example,
a massless spin-$s$ Lagrangian in $D+1$ dimensions KK reduces in $D$ dimensions to a theory containing symmetric tensor fields
of spin $0$ through $s$. The ($D+1$)-dimensional gauge invariance gives rise to the St\"{u}ckelberg symmetry in $D$ dimensions.
The tracelessness of the higher dimensional gauge parameter has non-trivial consequences; one can gauge away some, but not all,
of the auxiliary (St\"{u}ckelberg) fields. The gauge-fixed Lagrangian is then equivalent to the Singh-Hagen one up to field
redefinitions~\cite{ady}.

\section{Turning on Interactions}\label{sec:three}

As we already mentioned, turning on interactions for HS fields at the level of EoMs and constraints, by replacing ordinary
derivatives with covariant ones, results in inconsistencies. Let us take, for example, the na\"{i}ve covariant version of
Eqs.~(\ref{traceless})-(\ref{h2}) for massive spin 2 coupled to EM
\beq \left(D^2-m^2\right)\phi_{\mu\nu}=0,\qquad D\cdot\phi_\mu=0,\qquad \phi'=0,\eeq{kutta0}
where $D_\mu=\partial_\mu+ieA_\mu$. Now the Klein-Gordon equation and the transversality condition give
\beq \left[D^\mu,D^2-m^2\right]\phi_{\mu\nu}=0.\eeq{kutta1} The non-commutativity of covariant derivatives
then results in unwarranted constraints. For example, for a constant EM field strength $F_{\mu\nu}$, one obtains
\beq ieF^{\mu\rho}D_\mu\phi_{\rho\nu}=0.\eeq{kutta2} This constraint does not exist when the interaction is turned off,
so that the system of equations~(\ref{kutta0}) does not describe the same number of DoFs as the free theory.
Such difficulties can be avoided, as Fierz and Pauli suggested~\cite{pf}, by taking recourse to the Lagrangian
formulation. However, as we will see in Sections~\ref{sec:no-go} and~\ref{sec:vz}, the Lagrangian approach is not
free of difficulties either; inconsistencies show up for both massless and massive HS fields when interactions
are present.

\subsection{No-Go Theorems for Massless Fields}\label{sec:no-go}

No massless particle with $s>2$ has been observed in Nature. Neither is there any known string compactification
that gives a Minkowski space with massless particles of spin larger than two. Indeed, interactions of HS gauge
fields in flat space are severely constrained by powerful no-go theorems~\cite{w64,gpv,ad,ww,p}. Below we will closely
follow Ref.~\cite{p} to give an overview of some of the most important no-go theorems.

\subsubsection{Weinberg (1964)}

One can show, in a purely $S$-matrix-theoretic approach, that there are obstructions to consistent
long-range interactions mediated by massless bosons with $s>2$~\cite{w64}. Let us consider an $S$-matrix element with
$N$ external particles of momenta $p_i^\mu$, $i=1,...,N$ and a massless spin-$s$ particle of momentum $q^\mu$ .
The matrix element factorizes in the soft limit $q\rightarrow 0$ as:
\beq S(p_1,...,p_N,q,\epsilon)\rightarrow\sum_{i=1}^N\,g_i\,\left[\frac{p_i^{\mu_1}...
p_i^{\mu_s}\epsilon_{\mu_1..\mu_s}(q)}{2q\cdot p_i}\right]\,S(p_1,...,p_N),\eeq{2} where $\epsilon_{\mu_1..\mu_s}(q)$
is the polarization tensor of the spin-$s$ particle; it is transverse and traceless: \beq q^{\mu_1}\epsilon_{\mu_1
\mu_2...\mu_s}(q)=0, \qquad \epsilon'_{\mu_3..\mu_s}(q)=0. \eeq{3} The polarization tensor is redundant in that it
has more components than the physical polarizations of the massless particle. This redundancy is eliminated by demanding
that the matrix element vanish for spurious polarizations, i.e., for
\beq \epsilon^{(\text{spur})}_{\mu_1...\mu_s}(q)\equiv q_{(\mu_1}\alpha_{\mu_2..\mu_s)}(q),\eeq{4}
where $\alpha_{\mu_1...\mu_{s-1}}$ is also transverse and traceless. One finds from Eq.~(\ref{2}) that
the spurious polarizations decouple, for any generic momenta $p_i^\mu$, only if
\beq \sum_{i=1}^N g_i p_i^{\mu_1}...p_i^{\mu_{s-1}}=0.\eeq{5}
For $s=1$, this is possible when $\sum g_i=0$. This is nothing but the conservation of charge.
For $s=2$, Eq.~(\ref{5}) reduces to $\sum g_ip^\mu_i=0$, which can be satisfied for generic momenta if (i) $\sum p^\mu_i=0$,
and (ii) $g_i=\kappa$. The first condition imposes energy-momentum conservation, while the second one requires all
particles to interact with the same strength $\kappa$ with the massless spin 2 (graviton). The latter is simply the
equivalence principle. For $s\geq3$, Eq.~(\ref{5}) has no solution for generic momenta.

This argument does not rule out massless bosons with spin larger than 2, but say that they cannot give rise to
long-range interactions. Similar arguments were used in Ref.~\cite{gpv} for half-integer spins to forbid interacting
massless fermions with $s\geq\tfrac{5}{2}$. These theorems leave open the possibility that massless HS particles
may very well interact but can mediate only short-range interactions.

\subsubsection{Aragone-Deser (1979)}

In order to completely rule out massless HS particles, one should consider a truly universal interaction which no
particle can avoid. Gravitational coupling is one such interaction. In fact, Eq.~(\ref{5}) shows that the graviton
interacts universally with all matter in the soft limit $q\rightarrow 0$.

Aragone and Deser~\cite{ad} studied the gravitational coupling of a spin-$\tfrac{5}{2}$ gauge field to find that
such a theory is fraught with grave inconsistencies: the unphysical gauge modes decouple only in the free theory.
The system of a spin-$\tfrac{5}{2}$ field$-$described by the tensor-spinor $\psi_{\mu\nu}$$-$minimally coupled to gravity,
is governed by the action:
\beq S = -i\int d^Dx\sqrt{-g}\,\left[\bar{\psi}_{\mu\nu}\displaystyle{\not{\!\nabla}}\psi^{\mu\nu}
+2\bar{\displaystyle{\not{\!\!\psi}}}_\mu\displaystyle{\not{\!\nabla}}\displaystyle{\not{\!\!\psi}}^\mu
-\tfrac{1}{2}\bar{\psi}'\displaystyle{\not{\!\nabla}}\psi'+\left(\bar{\psi}'\nabla\cdot\displaystyle{
\not{\!\!\psi}}-2\bar{\displaystyle{\not{\!\!\psi}}}_\mu\nabla\cdot\psi^\mu-\text{h.c.}\right)\right].\eeq{6}
The redundancy in the field $\psi_{\mu\nu}$ should be eliminated, as in the free case, by the gauge invariance
\beq \delta\psi_{\mu\nu}=\nabla_{(\mu}\epsilon_{\nu)},\qquad\quad \gamma^\mu\epsilon_\mu=0.\eeq{8}
Under this gauge transformation, however, the action~(\ref{6}) transforms as \beq \delta S\sim\int d^D x\sqrt{-g}\left(i\bar{\epsilon}_\mu\gamma_\nu\psi_{\alpha\beta}R^{\mu\alpha\nu\beta}+\text{h.c.}\right),\eeq{9}
i.e., the action is invariant only in flat space where the Riemann tensor vanishes. In other words, the gauge modes
do not decouple for an interacting theory. It is easy to see that local non-minimal terms, regular in the neighborhood
of flat space, do not come to the rescue~\cite{ad}.

The Aragone-Deser no-go theorem is based on Lagrangian formalism, and therefore has a very important implicit assumption:
locality. Some non-locality in the Lagrangian appearing, for example, in the guise of a form factor for gravitational
couplings may remove the difficulties. It is also possible to have a non-minimal description involving a larger gauge
invariance than~(\ref{8}). All these issues can be addressed in the $S$-matrix language by considering, for example,
scattering of massless HS particles off soft gravitons.

\subsubsection{Weinberg-Witten (1980) \& Its Generalization}

Weinberg and Witten took the $S$-matrix approach to prove the following important theorem~\cite{ww}:
``A theory that allows the construction of a conserved Lorentz covariant energy-momentum tensor $\Theta^{\mu\nu}$ for which
$\int \Theta^{0\nu}d^3x$ is the energy-momentum four-vector cannot contain massless particles of spin $s>1$.''
They considered in 4D a particular matrix element: the elastic scattering of a massless
spin-$s$  particle off a soft graviton. The initial and final momenta of the spin-$s$ particle are $p$ and $p+q$
respectively, while the polarizations are identical, say $+s$. The off-shell graviton has a space-like momentum $q\rightarrow0$.
The matrix element is completely determined by the equivalence principle: \beq \lim_{q\rightarrow 0} \langle p+q, +s
| \Theta_{\mu\nu} | p, +s \rangle = p_\mu p_\nu, \eeq{11} where the normalization $\langle p| p'\rangle = (2\pi)^32p_0
\delta^3(\vec{p} -\vec{p}\,')$ has been used. For $q^2\neq0$, one can choose the ``brick wall''
frame~\cite{ww}, in which $p^\mu= (\,\tfrac{1}{2}|\vec{q}|,\tfrac{1}{2}\vec{q}\,)$
and $q^\mu= (0 , -\vec{q}\,)$, let us consider the effect of a rotation $R(\theta)$ by an angle $\theta$ around the $\vec{q}$
direction. The one-particle states transform as \bea R(\theta)|p,+s\rangle &=& \exp(\pm i\theta s)|p,+s\rangle, \label{12}\\
R(\theta)|p+q, +s\rangle &=& \exp(\mp i\theta s)|p+q,+s\rangle. \eea{13} The difference of the signs in the exponents arises
because $R(\theta)$ amounts to a rotation of $-\theta$ around $\vec{p} + \vec{q} = -\vec{p}$. Under spatial rotations,
$\Theta_{\mu\nu}$ decomposes into two real scalars, a vector and a symmetric traceless rank-2 tensor. These fields will be represented
by spherical tensors if we work in the basis in which the total angular momentum and projection thereof along the
$\vec{q}$-direction are the commuting variables. Combining the two real scalars into a complex one, we have the spherical tensors:
$\Theta_{0,0}$, $\Theta_{1,m}$ with $m=0,\pm 1$ and $\Theta_{2,m}$ with $m=0,\pm1,\pm2$. Now rotational invariance and the transformations
~(\ref{12})-(\ref{13}) give us \bea \langle p+q, +s| R^\dagger \Theta_{l,m} R | p, +s \rangle &=& e^{i\theta m}\langle p+q,
+s| \Theta_{l,m} | p,+s\rangle\nonumber\\ &=& e^{\pm 2i\theta s} \langle p+q, +s| \Theta_{l,m} | p, +s \rangle. \eea{14}
Since $|m|\leq 2$, the only solution to this equation for $s>1$ is that the matrix element must vanish. Because the helicities
are Lorentz invariant, the Lorentz covariance of $\Theta_{\mu\nu}$ implies, through Eq.~(\ref{14}), that the matrix element
vanishes in all frames as long as $q^2\neq0$. This is in direct contradiction with the equivalence principle as Eq.~(\ref{11})
gives a non-vanishing matrix element.

Note that the theorem does not apply to theories which do not have a Lorentz covariant energy-momentum tensor, e.g.,
supersymmetric theories and general relativity. In order to have a gauge invariant stress-energy tensor in these theories
one has to sacrifice manifest Lorentz covariance~\cite{dw2}. The same holds true for HS gauge fields. At any rate, the Weinberg-Witten
theorem can be extended to such theories by introducing unphysical states that correspond to spurious polarizations~\cite{p}.
Under Lorentz transformations, these unphysical polarizations mix with the physical ones such that the $\Theta_{\mu\nu}$
matrix elements are Lorentz covariant. The spurious states however must decouple from all physical matrix elements. By considering
one-graviton matrix elements one can show that for gravitationally interacting massless fields with $s>2$ decoupling of the
unphysical states is incompatible with the principle of equivalence~\cite{p}. Therefore, no massless HS particle with spin larger
than 2 can be consistently coupled to gravity in flat space. Similarly, one can show that no massless particles with
$s\geq\tfrac{3}{2}$ can have minimal EM coupling in flat space~\cite{p}.

\subsection{Pathologies of Interacting Massive Higher Spins}\label{sec:vz}

For massive HS particles, gauge transformation is no longer a symmetry of the action. So they are not subject to such algebraic
inconsistencies as their massless counterparts experience in the presence of interactions. However, they are plagued with
other pathologies as we will see below.

For simplicity, let us consider the massive spin-2 case. We begin by reviewing the key properties of the free theory, described
by the Fierz-Pauli Lagrangian~\cite{pf}:
\beq \mathcal L_{\text{FP}}=-\tfrac{1}{2}\,(\partial_\mu\vf_{\nu\rho})^2+(\partial_\mu\vf^{\mu\nu})^2 +\tfrac{1}{2}\,
(\partial_\mu\vf')^2\,-\partial_\mu\vf^{\mu\nu}\partial_\nu\vf'-\tfrac{1}{2}\, m^2[\vf_{\mu\nu}^2-\vf'^2].\eeq{f1}
Note that this Lagrangian can be obtained
either from the Lagrangian~(\ref{h13}) by the field redefinition:
$\varphi_{\mu\nu}=\phi_{\mu\nu}+\tfrac{1}{D-2}\,\eta_{\mu\nu}\phi$, that combines $\phi_{\mu\nu}$ and $\phi$ into
a single traceful field $\varphi_{\mu\nu}$, or from a Kaluza-Klein reduction of the Lagrangian~(\ref{h15}) with
a single dimension compactified on a circle of radius $1/m$. The corresponding EoMs read
\beq \mathcal R_{\mu\nu}\,\equiv\,(\Box-m^2)
\vf_{\mu\nu}-\eta_{\mu\nu}(\Box-m^2)\vf'+\partial_\mu\partial_\nu\vf'-\partial_{(\mu}\partial\cdot\vf_{\nu)}
+\eta_{\mu\nu}\partial\cdot\partial\cdot\vf_{\alpha\beta}=0.\eeq{f2}
Taking divergences and the trace of Eq.~(\ref{f2}) leads to
\bea &\partial^\mu \mathcal R_{\mu\nu}\equiv-m^2(\partial\cdot\vf_{\nu}-\partial_\nu\vf')=0,&\label{f3}\\
&\partial^\mu\partial^\nu\mathcal R_{\mu\nu}\equiv-m^2(\partial\cdot\partial\cdot\vf-\Box\vf')=0,&\label{f4}\\
&\mathcal R^\mu_{~\mu}\equiv(D-2)(\partial\cdot\partial\cdot\vf-\Box\vf')+[(D-1)m^2]\,\vf'=0.&\eea{f5}
Combining Eqs.~(\ref{f4}) and (\ref{f5}) one arrives at an interesting consequence,
\beq [(D-1)m^2]\vf'=0.\eeq{f6}
Thus, for $m^2\neq0$ and $D>1$ one is led to the dynamical trace constraint:
\beq \vf~=~0~.\eeq{f7}
Then from Eq.~(\ref{f3}) one gets the transversality condition:
\beq \partial\cdot\phi_\mu=0,\eeq{doublekutta}
so that finally Eq.~(\ref{f2}) reduces to the Klein-Gordon equation
\beq (\Box-m^2)\vf_{\mu\nu}=0,\eeq{f9}
which is manifestly hyperbolic and causal. Eqs.~(\ref{f7})-(\ref{f9}) are the simplest instance of a Fierz--Pauli system,
which originates from the Lagrangian~(\ref{f1}) in arbitrary space-time dimensions.

Now let us complexify the spin-2 field in the Lagrangian~(\ref{f1}) in order to minimally couple it to a constant EM
background. The minimal coupling is ambiguous because covariant derivatives do not commute, so that we are actually led to
a family of Lagrangians containing one parameter, which we call the gyromagnetic ratio $g$ (see, e.g.,~\cite{deser}):
\bea \mathcal L&=&-|D_\mu\vf_
{\nu\rho}|^2+2|D\cdot\varphi^{\mu}|^2+|D_\mu \varphi'|^2+(\varphi^*_{\mu\nu} D^\mu D^\nu \vf'+\text{c.c.})
-m^2(\vf^*_{\mu\nu}\vf^{\mu\nu}-\vf'^*\vf')\nonumber\\&&+\,2ieg\,\text{Tr}(\vf\cdot F\cdot\vf^*).\eea{vz1}
The resulting EoMs are
\beq 0=\mathcal R_{\mu\nu}\equiv(D^2-m^2)\left(\vf_{\mu\nu}-\eta_{\mu\nu}\vf'\right)+\tfrac12 D_{(\mu} D_{\nu)}\vf'-D_{(\m}
D\cdot\vf_{\n)}+\eta_{\mu\nu}D\cdot D\cdot\vf+iegF_{\r(\m}\vf_{\n)}^{~\r}.\eeq{vz2}
Combining the trace and the double divergence of Eq.~\eqref{vz2} gives
\beq \left(\tfrac{D-1}{D-2}\right)m^4\vf'=ie(2g-1)F^{\m\n}D_\m D\cdot\vf_{\n}+(g-2)e^2F^{\m\r}F_\r^{~\n}
\vf_{\m\n}-\tfrac34\,e^2F^{\m\n}F_{\m\n}\vf'.\eeq{vz6}
The first term on the right-hand side signals a potential breakdown of the DoF count, since the divergence constraint of the
free theory is turned into a propagating field equation unless $g=\12$. The unique minimally coupled model that does not
give rise to a wrong DoF count has therefore $g=\12$, and is precisely the Federbush Lagrangian of~\cite{fed}. With this choice,
one obtains
\bea &D\cdot\vf_{\n}-D_\n\vf'=\tfrac32
(ie/m^2)\left[F^{\r\s}D_\r\vf_{\s\n}-F_{\n\r}D\cdot\vf^{\r}+F_{\n\r}D^\r\vf'\right],&\label{vz7}\\&D\cdot D\cdot
\vf-D^2\vf'=\tfrac32(1/m^2)\left[\,\text{Tr}(F\cdot\vf\cdot F)-\tfrac12\text{Tr}F^2\vf'\,\right]~,&
\label{vz8}\\&\vf'=-\tfrac32\left(\tfrac{D-2}{D-1}\right)(e/m^2)^2\left[\,\text{Tr}(F\cdot\vf\cdot F)
-\tfrac12\text{Tr}F^2\vf'\,\right].&\eea{vz9}
The trace constraint can also be rewritten as
\beq \vf'=-\frac{\tfrac32\left(\tfrac{D-2}{D-1}\right)(e/m^2)^2\,\text{Tr}(F\cdot\vf\cdot F)}
{1-\tfrac34\left(\tfrac{D-2}{D-1}\right)(e/m^2)^2\,\text{Tr}F^2}.\eeq{vz10}

Unlike in the free theory, the trace does not vanish in the presence of the EM background. This expression, however,
makes perfect sense for small values of the EM field invariants in units of $m^2/e$\,. If some invariant were ${\cal O}(1)$
in those units, a number of new phenomena, including Schwinger pair production~\cite{schwinger} and Nielsen-Olesen
instabilities~\cite{nielsen}, would appear. The very existence of these instabilities implies that any effective Lagrangian
for a charged particle interacting with EM can be reliable, even well below its own cutoff scale, only for small EM field invariants.

One still needs to see whether the dynamical DoFs propagate in the correct number and causally. To this end, let us isolate
the terms in Eqs.~(\ref{vz2}) that contain second-order derivatives:
\beq \mathcal R_{\mu\nu}^{(2)}=D^2\vf_{\mu\nu}-\left[D_\m(D\cdot\vf_{\n}-D_\n\vf)+
(\m\leftrightarrow\n)\right]-\tfrac12 D_{(\mu} D_{\nu)}\vf+\eta_{\mu\nu}(D\cdot D\cdot\vf-D^2\vf').\eeq{vz11}
The last term can be actually dropped in view of~(\ref{vz8}), while the constraint equations~(\ref{vz7}) and~(\ref{vz9})
can be substituted in the second and third terms to obtain
\bea \mathcal R_{\mu\nu}^{(2)}&=&\Box\vf_{\m\n}-\tfrac{3}{2}(ie/m^2)\left[F^{\r\s}\de_\r\de_{(\m}\vf_{\n)\s}
+F_{\r(\m}\de_{\n)}(\de\cdot\vf^{\r}-\de^\r\vf')\right]\nn\\&&+\tfrac32\left(\tfrac{D-2}{D-1}\right)(e/m^2)^2
\left[F^{\r\s}F_\s^{~\l}\de_\m\de_\n\vf_{\r\l}-\tfrac12\text{Tr}F^2\de_\m\de_\n\vf'\right].\eea{vz12}
Following~\cite{vz} one can employ the method of characteristic determinant to investigate the causal properties of
the system. One replaces $i\partial_\mu$ with $n_\mu$, the normal to the characteristic hypersurfaces, in the highest-derivative
terms of the EoMs\footnote{This procedure is akin to solving the EoMs in the eikonal approximation letting
$\vf_{\m\n}=\hat{\vf}_{\m\n}\exp(itn\cdot x),\,t\rightarrow\infty$.}. The determinant $\Delta(n)$ of the resulting
coefficient matrix determines the causal properties of the system:
if the algebraic equation $\Delta(n)=0$ has real solutions for $n_0$ for any $\vec{n}$, the system is hyperbolic, with maximum
wave speed $n_0/|\vec{n}|$. On the other hand, if there are time-like solutions $n_\mu$ for $\Delta(n)=0$, the system admits
faster-than-light propagation. The coefficient matrix appearing in~(\ref{vz12}) takes the form
\bea M_{(\m\n)}^{~~(\a\b)}(n)&=&-\tfrac12n^2
\delta^{(\a}_\m\delta^{\b)}_\n+\tfrac34(ie/m^2)\left[n_\r F^{\r(\a}n_{(\m}\delta_{\n)}^{\b)}-n_{(\m}F_{\n)}^{~(\a}n^{\b)}
+2n_{(\m}F_{\n)}^{~\r}n_{\r}\eta^{\a\b}\right]\nn\\&&-\,\tfrac32\left(\tfrac{D-2}{D-1}\right)(e/m^2)^2\,n_\m n_\n\left[F^{\a\r}F_\r^{~\b}-\tfrac12\text{Tr}
F^2\eta^{\a\b}\right]~.\eea{vz13}
Note that this expression is to be regarded as a matrix whose $\tfrac12D(D+1)$ rows and columns are labeled by pairs
of Lorentz indices $(\m\n)$ and $(\a\b)$. In particular, in $D=4$ its determinant reads
\beq \Delta(n)=(n^2)^8\left[n^2-\left(\tfrac{e}{m^{2}}\right)^2\left(\tilde{F}\cdot n\right)^2\right]
\left[n^2+\left(\tfrac{3e}{2m^{2}}\right)^2\left(\tilde{F}\cdot n\right)^2\right],\eeq{vz14}
where $\tilde{F}_{\m\n}\equiv\tfrac12\e_{\m\n\r\s}F^{\r\s}$, so that
$(\tilde{F}\cdot n)^2\equiv(n_0\vec B+\vec n\times\vec E)^2-(\vec n\cdot\vec B)^2$.

Let us now consider 4D EM field invariants such that $\vec B\cdot\vec E=0$, $\vec B^2-\vec E^2>0$,
which entails that $\vec B^2$ is non-vanishing in all Lorentz frames. One can always find a vector $\vec n$,
perpendicular to $\vec B$, for which the characteristic determinant vanishes provided that
\beq \frac{n_0}{|\vec n|}=\frac{1}{\sqrt{1-\left(\tfrac{3e}{2m^{2}}\right)^2\vec B^2}}\,,\eeq{vz15}
thanks to the last factor appearing in~(\ref{vz14}).
Therefore, superluminal propagation can take place even for infinitesimally small values of $\vec B^2$. Moreover, the
propagation itself ceases to occur whenever $\vec B^2\geq\left(\tfrac{2}{3}m^2/e\right)^2$. One can always find a
Lorentz frame where the pathology shows up: the magnetic field $\vec B^2$ can reach the critical value in
a frame where $\vec E^2=\left(\tfrac{2}{3}m^2/e\right)^2-\e$, with $\e$ arbitrarily small.
This is the most serious aspect of the problem: it persists even for infinitesimally small values of the EM field
invariants$-$far away from the instabilities~\cite{schwinger,nielsen}$-$so that one expects to have well-behaved
and long-lived propagating particles. This is the so-called Velo-Zwanziger problem~\cite{vz}, which as we see
arises already at the classical level\footnote{The pathology in the corresponding quantum theory was found much
earlier by Johnson and Sudarshan~\cite{js} for massive spin $\tfrac{3}{2}$. From a canonical point of view,
in an EM background, the equal time commutation relations become ill-defined. That the Johnson-Sudarshan and
Velo-Zwanziger problems have a common origin was later shown in~\cite{same}.}.

Note that the pathology is not a special property of the spin-2 case; it generically shows up for all charged massive
HS particles with $s\geq3/2$ since it originates from the very existence of longitudinal modes for massive HS
particles~\cite{pr1}. More importantly, it persists for a wide class of non-minimal generalizations of the
theory~\cite{kob,d3,sham}. Thus, field theoretically it is quite challenging to construct consistent interactions
of massive HS particles.

\section{Remedy for the Velo-Zwanziger Problem}\label{sec:open}

The superluminal propagation of the physical modes of a massive HS field in an external field is a serious problem.
However, addition of non-minimal terms and/or new dynamical DoFs may come to the rescue. For example, the Lagrangian
proposed in Ref.~\cite{PR2} incorporates appropriate non-minimal terms that propagate only within the light cone
the physical modes of a massive spin-$\tfrac{3}{2}$ field in a constant external EM background. A more well-known example is
$\mathcal{N}=2$ (broken) SUGRA~\cite{SUGRA1,SUGRA2}: it contains a massive gravitino that propagates consistently,
even when the cosmological constant vanishes, if it has a charge, $e=\tfrac{1}{\sqrt{2}}(m/M_\text{P})$, under the
graviphoton~\cite{DZ}. Causality is preserved by the presence of $both$ EM and gravity, along with non-minimal terms.

In this Section we will see how String Theory bypasses the Velo-Zwanziger problem. To this end, one starts with
the (exactly solvable) $\sigma$-model for bosonic open string in a constant EM background, originally considered
in~\cite{Abouelsaood}, and perform a careful analysis of the mode expansion and the Virasoro generators.
The physical state conditions for string states can be translated into the language of string fields to obtain
a Fierz-Pauli system, which explicitly shows that indeed the Velo-Zwanziger problem is absent for the symmetric
tensors of the first Regge trajectory~\cite{AN2,PRS}.

\subsection{Open Strings in Constant EM Background}\label{sec:sigma}

We consider an open bosonic string whose endpoints lie on a space-filling $D$-brane\footnote{We set the string tension
to $\alpha'=\tfrac{1}{2}$.}. A Maxwell field $A_\mu$ living in
the world-volume of the $D$-brane couples to charges $e_0$ and $e_\pi$ at the string endpoints. For an electromagnetic
background with constant field strength $F_{\mu\nu}$, the EoMs are those of the usual free strings while the boundary
conditions are deformed, so that the $\sigma$-model is exactly solvable~\cite{Abouelsaood,AN2,AN1}. With a careful
definition of the mode expansion, it is possible to have commuting center-of-mass coordinates, which obey
canonical commutation relations with the (non-commuting) covariant momenta~\cite{AN2,PRS}. Like for free strings,
there is an infinite set of creation and annihilation operators that are well-defined in the physically interesting
regimes~\cite{AN2,PRS}.

An important point is that the mode $\alpha_0^\mu$ is the covariant momentum only up to a rotation:
\beq \alpha_0^\mu=\left(\sqrt{G/eF}\right)^\m_{~~\n}\,p^\nu_{\text{cov}},\eeq{r29}
where $e=e_0+e_\pi$ is the total charge of the string, and
\beq G=\frac{1}{\pi}\left[\,\tanh^{-1}(\pi e_0F)+\tanh^{-1}(\pi e_\p F)\,\right].\eeq{r13}
With $\left[p^\mu_{\text{cov}},p^\nu_{\text{cov}}\right]=ieF^{\mu\nu}$, it is convenient to define a different covariant
derivative:
\beq \mathcal D^\m\equiv i\alpha_0^\m,\qquad\quad [\mathcal D^\m,\mathcal D^\n]=-iG^{\m\n}.\eeq{fancyD}

The commutation relations obeyed by the Virasoro generators can be worked out as usual. The end result is the
emergence of an additive contribution to $L_0$~\cite{Abouelsaood}:
\beq L_0\rightarrow L_0+\tfrac{1}{4}\text{Tr}G^2,\eeq{r52}
Up to this shift, the Virasoro algebra remains precisely the same as in the free theory:
\beq [L_m,L_n]=(m-n)L_{m+n}+\tfrac{1}{12}\,D(m^3-m)\delta_{m,-n},\eeq{r53}
The shift~(\ref{r52}), however, has an important effect since it deforms the masses of the open-string excitations.
%

\subsection{Physical State Conditions}\label{sec:PSCEM}

A generic string state in the presence of a constant EM background is constructed the same way as in the free theory,
since the two cases have identical sets of creation and annihilation operators. It is given by:
\beq |\Psi\rangle=\sum_{s=1}^\infty\sum_{m_i=1}^\infty \psi_{\mu_1...\mu_s}^{(m_1...m_s)}(x)\,a^{\dag\mu_1}_{m_1}\cdots
a^{\dag\mu_s}_{m_s}\,|0\rangle,\eeq{kuttarbaccha} where the rank-$s$ coefficient tensor is interpreted as a string field,
and as such is a function of the string center-of-mass coordinates.

Because the Virasoro algebra is the same as in the free theory, a string state $|\Phi\rangle$ is called ``physical'' if
it satisfies the conditions~\cite{Abouelsaood,AN2,AN1}:
\bea (L_0-1)|\Phi\rangle&=&0,\label{b11}\\ L_1|\Phi\rangle
&=&0,\label{b12}\\L_2|\Phi\rangle&=&0,\eea{b13}
where now
\bea L_0&=&-\tfrac{1}{2}\mathcal{D}^2+\sum_{m=1}^{\infty}(m+iG)_{\mu\nu}\,a_m^{\dag\mu}a_m^\nu+
\tfrac{1}{4}\,\text{Tr}G^2\nonumber\\&\equiv&-\tfrac{1}{2}\mathcal{D}^2+\left(\mathcal{N}+\tfrac{1}{4}\,
\text{Tr}G^2\right)+i\sum_{m=1}^{\infty}G_{\mu\nu}\,a_m^{\dag\mu}a_m^\nu~,\label{b9.1}\\L_1&=&-i
\left[\sqrt{1+iG}\right]_{\mu\nu}\mathcal{D}^\mu a_1^\nu+\sum_{m=2}^{\infty}\left[\sqrt{(m+iG)
(m-1+iG)}\right]_{\mu\nu}a_{m-1}^{\dag\mu}a_m^\nu~,\eea{b9.2} \bea L_2~~=~~-i\left[\sqrt{2+iG}\right]
_{\mu\nu}\mathcal{D}^\mu a_2^\nu&+&\tfrac{1}{2}\left[\sqrt{1+G^2}\right]_{\mu\nu}a_1^\mu a_1^\nu
\nonumber\\&+&\sum_{m=3}^{\infty}\left[\sqrt{(m+iG)(m-2+iG)}\right]_{\mu\nu}a_{m-2}^{\dag\mu}
a_m^\nu~,\eea{b9.3}
with $\mathcal D_\m$ defined in Eq.~(\ref{fancyD}). Notice that the number operator $\mathcal{N}$ will
be defined in such a way that its eigenvalues $N$ are integers, just as for free strings:
\beq \mathcal{N}\equiv\sum_{n=1}^\infty n \, a_n^\dag\cdot a_n.\eeq{b7}
We would like to see how the Eqs.~(\ref{b11})-(\ref{b13}) translate into the string field theory language.

\subsubsection{Level $N=1$}

The generic state at this level is given by
\beq |\Phi\rangle=A_\mu(x)\,a^{\dag\mu}_1\,|0\rangle.\eeq{b16}
While Eq.~(\ref{b13}) is empty, Eqs.~(\ref{b11}) and~(\ref{b12}) give
\bea &\left[\mathcal{D}^2-\tfrac{1}{2}\text{Tr} G^2\right]A_\mu-2iG_\mu^{~\nu}A_\nu=0,&\label{b17}\\
&\mathcal{D}^\mu\left(\sqrt{\mathbf{1}+iG}\cdot A\right)_\mu=0,&\eea{b18}
thanks the commutation relations of the creation and annihilation operators, and the usual definition
of the oscillator vacuum. By the field redefinition,
\beq \mathcal{A}_\mu\equiv\left(\sqrt{\mathbf{1}+iG}\cdot A\right)_\mu,
\eeq{def1}
the field equations~(\ref{b17})-(\ref{b18}) can be cast in the form
\bea &\left[\mathcal{D}^2-
\tfrac{1}{2}\text{Tr}G^2\right]\mathcal{A}_\mu-2iG_\mu^{~\nu}\mathcal{A}_\nu=0,&\label{b19}\\
&\mathcal{D}^\mu\mathcal{A}_\mu=0.&\eea{b20}
The algebraic consistency of this system can be easily verified. The divergence constraint~(\ref{b20})
ensures that the number of dynamical DoFs is not affected.

\subsubsection{Level $N=2$}

A generic state at this mass level is written as
\beq |\Phi\rangle=h_{\mu\nu}(x)\,a^{\dag\mu}_{1}a^{\dag\nu}_{1}\,
|0\rangle+\sqrt{2}iB_\mu(x)\,a^{\dag\mu}_{2}\,|0\rangle.\eeq{b21}
After the field redefinitions
\bea
\mathcal{H}_{\mu\nu}&\equiv&\left(\sqrt{1+iG}\cdot h\cdot\sqrt{1-iG}\,\right)_{\mu\nu},\label{def2}\\
\mathcal{B}_\mu&\equiv&\left(\sqrt{1+\tfrac{i}{2}G}\cdot B\right)_\mu,\eea{def3}
the physical state conditions~(\ref{b11})-(\ref{b13}) leave us with
\bea &\left(\mathcal{D}^2-2-\tfrac{1}{2}\,\text{Tr}G^2\right)\mathcal{H}_{\mu\nu}-2i(G\mathcal{H}
-\mathcal{H}G)_{\mu\nu}=0,&\label{b25}\\&\left(\mathcal{D}^2-2-\tfrac{1}{2}\,\text{Tr}G^2\right)
\mathcal{B}_\mu-2iG_\mu^{~\nu}\mathcal{B}_\nu=0,&\label{b26}\\&\mathcal{D}^\mu\mathcal{H}_{\mu\nu}
-(1+iG)_\nu^{~\rho}\mathcal{B}_\rho=0,&\label{b27}\\&\mathcal{H}^\mu_{~\mu}+2\mathcal{D}^\mu
\mathcal{B}_\mu=0.&\eea{b28}
One finds that the system~(\ref{b25})-(\ref{b28}) enjoys an on-shell gauge invariance with a vector gauge parameter
in $D=26$~\cite{PRS}. Thus, in $D=26$, one can gauge away the vector field $\mathcal B_\mu$ to obtain
\bea &\left(\mathcal{D}^2-2-\tfrac{1}{2}\,\text{Tr}G^2\right)\mathcal{H}_{\mu\nu}-2i(G\cdot\mathcal{H}
-\mathcal{H}\cdot G)_{\mu\nu}=0,&\label{b37.1}\\&\mathcal D\cdot\mathcal H_{\mu}=0,&\label{b37.2}\\
&\mathcal H'=0.&\eea{b37.3}
Thus $\mathcal H_{\mu\nu}$ is a massive spin-2 field, with $\text{mass}^2=(1/\alpha')\left(1+\tfrac{1}{4}
\text{Tr}G^2\right)$, that possesses a suitable non-minimal coupling to the background EM field. It is easy to see
that the system~(\ref{b37.1})-(\ref{b37.3}) is algebraically consistent and preserves the right number of DoFs,
namely $\12(D+1)(D-2)$.

\subsubsection{Arbitrary mass level: $N=s$}\label{sec:arbitrary}

As one goes higher in the mass level, subleading Regge trajectories show up. It turns out that the lower Regge trajectories
are not consistent in isolation, whereas the leading Regge trajectory is~\cite{PRS}. At level $N=s$, the first Regge trajectory
contains a symmetric rank-$s$ tensor. The string field equations that result from the physical state
conditions~\eqref{b11}-\eqref{b13} are~\cite{PRS}:
\bea &\left[\mathcal{D}^2-2(s-1)-\tfrac{1}{2}\,\text{Tr}G^2\right]\varPhi_{\mu_1...\mu_s}+2iG^\alpha{}_{~(\mu_1}
\varPhi_{\mu_2...\mu_s)\alpha}=0,&\label{b60}\\&\mathcal{D}\cdot\varPhi_{\mu_2...\mu_s}=0,&\label{b61}\\
&\varPhi'_{\mu_3...\mu_s}=0.&\eea{b62}
One can easily show that the Eqs.~(\ref{b60})-(\ref{b62}) are algebraically consistent. They form a Fierz-Pauli system for
a massive spin-$s$ field, with $\text{mass}^2=(1/\alpha')\left(s-1+\tfrac{1}{4}\text{Tr}G^2\right)$.
It is manifest that the system gives the correct count of DoFs. Following~\cite{AN2,PRS}, we will now show that these
equations indeed propagate the physical DoFs within the light cone.

\subsection{Proof of Causal Propagation}\label{sec:causal}

We employ the method of characteristic determinants, already discussed in Section~\ref{sec:vz}. Let us note from
Eqs.~(\ref{b60})-(\ref{b62}) that the highest-derivative terms appearing in the EoMs boil down to the scalar operator
$\mathcal D^2$ acting on the field. From the definition~(\ref{fancyD}) of $\mathcal D_\m$, it is clear that the
vanishing of the characteristic determinant boils down to the condition:
\beq \left(G/eF\right)^\m_{~~\n}\ n_\m \, n^\n=0.\eeq{c1}
One can go to a Lorentz frame in which $F$ is block skew-diagonal:
$F^{\m}_{~~\n}=\text{diag}\left(\,F_1~,F_2~,F_3~,...~...\,\right)$,
with the blocks given by
\beq F_1=a\left(\begin{array}{cc}
               0 & 1 \\
               1 & 0 \\
             \end{array}
           \right),\qquad F_{i\neq1}=b_i\left(
             \begin{array}{cc}
               0 & 1 \\
               -1 & 0 \\
             \end{array}
           \right),
\eeq{c3}
where $a$ and $b_i$'s are real-valued functions of the EM field invariants whose values are always small in physically interesting
cases. In the same Lorentz frame, clearly $G$ will be block skew-diagonal
as well: $G^{\m}_{~\n}=\text{diag}\left(\,G_1~,G_2~,G_3~,...~...\,\right)$, with
\beq G_1=f(a)\left(
             \begin{array}{cc}
               0 & 1 \\
               1 & 0 \\
             \end{array}
           \right),\qquad G_{i\neq1}=g(b_i)\left(
             \begin{array}{cc}
               0 & 1 \\
               -1 & 0 \\
             \end{array}
           \right),\eeq{c4}
where the functions $f$ and $g$ are given as
\bea f(a)&\equiv& \frac{1}{\pi}\, [\,\tanh^{-1}(\pi e_0a)+
\tanh^{-1}(\pi e_\pi a)\,],\label{c5}\\ g(b_i)&\equiv&\frac{1}{\pi}\, [\,\tan^{-1}(\pi e_0b_i)+
\tan^{-1}(\pi e_\pi b_i)\,].\eea{c6}

We emphasize that if the EM field invariants are small, these functions are always well-defined and their absolute
values are much smaller than unity. Given the forms~(\ref{c3}) and~(\ref{c4}), one finds that ($G/eF$) is in fact a diagonal matrix:
\beq \left(\frac{G}{eF}\right)^\m_{~\n}=\text{diag}\left[\,\frac{f(a)}{ea}~,
\frac{f(a)}{ea}~,\frac{g(b_2)}{eb_2}~,\frac{g(b_2)}{eb_2}~,\frac{g(b_3)}{eb_3}~,\frac{g(b_3)}{eb_3}~,
...~...\,\right].\eeq{c7}
The functions~(\ref{c5}) and (\ref{c6}), however, satisfy the inequalities
\beq \frac{f(a)}{ea}\geq 1,\qquad 0<\frac{g(b_i)}{eb_i}\leq 1.\eeq{c8}
Then, in view of~(\ref{c7}), any solution $n_\mu$ of  Eq.~(\ref{c1}) must be space-like:
\beq n^2\geq0,\eeq{c9}
which is a Lorentz invariant statement. We therefore conclude that the propagation of the field is causal
in all Lorentz frames.

We finish this Section with a few remarks. We have seen that String Theory gives a consistent set of EoMs
and constraints~(\ref{b60})-(\ref{b62}) so that any field belonging to the first Regge trajectory propagates
causally in a constant EM background. It is also guaranteed that this system comes from a Lagrangian~\cite{PRS}.
In fact, explicit Lagrangians have been worked out for $s=2$~\cite{AN2} and $s=3$~\cite{string3}. They, however,
give rise to the Fierz-Pauli conditions~(\ref{b60})-(\ref{b62}) only in the critical dimension $D=26$~\cite{PRS}.
To obtain a consistent theory in $D<26$, one may perform a consistent dimensional reduction of the string-theory
Lagrangian by keeping only the singlets of the internal coordinates. In this way, one may start with the
spin-2 Lagrangian~\cite{AN2} and ends up having a consistent model, say in 4D, that contains spin-2 plus a spin-0
field both having the same mass and charge~\cite{2plus0}.

Note that the generalized Fierz-Pauli conditions~(\ref{b60})-(\ref{b62}) contain non-standard kinetic contributions.
One may wonder whether the flat-space no-ghost theorem extends to this case. One can show that indeed the no-ghost
theorem continues to hold in the regime of physical interest~\cite{PRS}.

Last but not the least, let us notice that Open String Theory requires a universal value of $g=2$ for the
gyromagnetic ratio for all spin~\cite{Ferrara}. This value can simply be read from the non-minimal term appearing
in Eq.~(\ref{b60}). At the Lagrangian level, this can be seen by removing the kinetic-like cubic interactions by
suitable field redefinitions, which uniquely give $g=2$~\cite{PRS}.

\section{Intrinsic Cutoff for Interacting Massive Higher Spins}\label{sec:five}

Can an interacting massive HS particle be described by a local Lagrangian up to arbitrarily high energy scales?
The answer is no, because the no-go theorems~\cite{w64,gpv,ad,ww,p} on EM and gravitationally coupled massless
particles imply that the cutoff of that Lagrangian must vanish in the massless limit. One would like to find the
explicit dependence of the cutoff on the mass and the coupling constant of the theory to see if the cutoff can be
parametrically larger than the mass.

The existence of a finite cutoff and the origin of the acausality problem are both due to a simple fact:
the free kinetic term of a HS field exhibits gauge invariance, i.e., it has zero modes. In the presence of
interactions, some of these modes may acquire non-canonical kinetic term and propagate acausally. On the
other hand, in the massless limit the HS propagator is singular, so that scattering amplitudes diverge.
All this is not manifest at all in the unitary gauge. The best way to understand these issues is the
St\"uckelberg formalism, which focuses precisely on the gauge modes. To understand this formalism, let us
consider again the Fierz-Pauli Lagrangian~(\ref{f1}) for a massive spin-2 field\footnote{For this field,
the Stu\"ckelberg formalism has been employed, for example, in~\cite{pr1,Nima,Claudia}.}. Note that gauge
invariance can be restored in the massive theory by introducing the Stu\"ckelberg fields $B_\mu$ and $\phi$,
through the field redefinition:
\beq \vf_{\m\n}\rightarrow\tilde\vf_{\m\n}=\vf_{\m\n}+\frac{1}{m}\,\de_{\mu}\left(B_\nu-\frac{1}{2m}
\de_{\nu}\phi\right)+\frac{1}{m}\,\de_{\nu}\left(B_\mu-\frac{1}{2m}\de_{\mu}\phi\right),\eeq{sbrg1}
so that the following transformations become a gauge symmetry:
\bea \delta\vf_{\mu\nu} &=&\de_{(\mu}\lambda_{\nu)},\label{sbrg3}\\
\delta B_{\mu} &=& \de_{\mu}\lambda - m\lambda_{\mu},\label{sbrg3.5}\\
\delta\phi &=& 2m\lambda.\eea{sbrg4}
Note that this St\"uckelberg symmetry is a fake one in that one can always gauge away the St\"uckelberg
fields to go back to the unitary-gauge Fierz-Pauli Lagrangian~(\ref{f1}). However, these fields may help us
to understand some otherwise obscure phenomena. For example, one slick way of understanding the mass
term, $m^2\left(\vf^{\m\n}\vf_{\m\n}-\vf'^2\right)$, in~(\ref{f1}) is that this is the unique linear
combination of $\vf^{\m\n}\vf_{\m\n}$ and $\vf'^2$ for which 4-derivative bad kinetic
terms of the field $\phi$ are killed.

One can instead choose a gauge fixing such that all the fields acquire canonical kinetic terms. In the
interacting theory, one thus ends up having non-renormalizable interactions\footnote{In this Section,
we will work only in $D=4$.} involving the St\"uckelberg fields, weighted by coupling constants with
negative mass dimensions~\cite{pr1,Nima,pr2}:
\beq \mathcal L=\mathcal L_{\text{renormalizable}}+\sum_{n>0}(\Lambda_n)^{-n}\mathcal O_{n+4},\eeq{sbrg5}
where $\mathcal O_{n+4}$ denotes operators of dimension $n+4$. Some of these operators may be eliminated
by field redefinitions or by adding non-minimal terms to the Lagrangian. The smallest $\Lambda_n$ in the
remaining terms defines the ultimate cutoff of the effective field theory. Let us now consider some examples:
massive particles coupled to EM in flat space.

\subsection*{Spin 1}

We start with a complex massive spin-1 field, $W_\mu$, minimally coupled to EM:
\beq \mathcal L=-\tfrac{1}{2}\,|D_\mu W_\nu -D_\nu W_\mu|^2 -m^2W_\mu^*W^\mu-\tfrac{1}{4}F_{\m\n}^2,\eeq{o1}
Contrary to na\"ive power counting, we know that this Lagrangian is non-renormalizable. To make the
higher dimensional operators appear, we introduce a scalar $\phi$ through the substitution
\beq W_\mu=V_\mu-\frac{1}{m}\,D_\mu \phi.\eeq{ooo1} so that one has the gauged St\"uckelberg symmetry:
\bea \delta V_\mu&=&D_{\mu}\lambda,\\ \delta\phi &=& m\lambda.\eea{o2} With the addition of the gauge-fixing
term \beq \mathcal L_{\text{gf}}=-|D\cdot V-m\phi|^2,\eeq{ooo3} one is left with the following operators:
\bea \mathcal L+\mathcal L_{\text{gf}}&=&V^*_\m\left(D^2-m^2\right)V^\mu+\phi^*\left(D^2-m^2\right)\phi
-\tfrac{1}{4}F_{\m\n}^2\nn\\&&-\left[\frac{ie}{2m}\,F^{\mu\nu} \phi^*(D_\mu V_\nu-D_\nu V_\mu) +\text{c.c.}\right]
-\frac{e^2}{2m^2}\,F_{\mu\nu}^2\phi^*\phi.\eea{o3}
The presence of non-renormalizable interactions implies the existence of a UV cutoff $\Lambda\sim m/e$. In a local
theory addition of non-minimal terms and field redefinitions cannot remove these operators without introducing new
ones that instead lower the cutoff, parametrically in $e\ll1$~\cite{pr1,pr2}.

Thus the cutoff of a charged massive spin-1 particle is $\Lambda\sim m/e$ in the absence of other DoFs. This means
in particular that the electrodynamics of $W^\pm$ boson has a cutoff $\Lambda\sim 270\,\text{GeV}$. Below this scale
some new physics must appear. This could be strong coupling or new dynamical DoFs. We now know that the latter
possibility is realized in nature. The new DoF$-$the neutral Higgs scalar$-$has a mass of $126\,\text{GeV}$, which
is indeed below the cutoff scale $\Lambda$.

\subsection*{Spin 2}

We take the Pauli-Fierz Lagrangian~(\ref{f1}), introduce the St\"{u}ckelberg fields through~(\ref{sbrg1}), make the
fields complex, and then replace ordinary derivatives with covariant ones. Thus result is
\beq \mathcal L=-|D_\mu\tilde{\vf}_{\nu\rho}|^2+2|D_\mu\tilde{\vf}^{\mu\nu}|^2 +|D_\mu\tilde{\vf}'|^2-[D_\mu\tilde{\vf}
^{*\mu\nu}D_\nu\tilde{\vf}'+\text{c.c.}]-m^2[\tilde{\vf}_{\mu\nu}^*\tilde{\vf}^{\mu\nu}-\tilde{\vf}'^*\tilde{\vf}'],\eeq{rr7}
with
\beq \tilde{\vf}_{\mu\nu}=\vf_{\mu\nu}+\frac{1}{m}\,D_{\mu}\left(B_\nu-\frac{1}{2m}D_{\nu}\phi\right)+\frac{1}{m}\,D_{\nu}
\left(B_\mu-\frac{1}{2m}D_{\mu}\phi\right).\eeq{rr8}
This system enjoys the covariant St\"{u}ckelberg symmetry:
\bea \delta\vf_{\mu\nu} &=&D_{(\mu}\lambda_{\nu)},\label{rr9}\\
\delta B_{\mu} &=& D_{\mu}\lambda - m\lambda_{\mu},\label{rr10}\\
\delta\phi &=& 2m\lambda.\eea{rr11}

Now we diagonalize the kinetic operators and thereby make sure that the propagators are smooth in the massless limit.
This is done by the field redefinition:
\beq \vf_{\mu\nu} \rightarrow \vf_{\mu\nu}- \tfrac{1}{2}\,\eta_{\mu\nu}\phi,\eeq{rr14}
along with the addition of the gauge-fixing terms (thereby exhausting all gauge freedom):
\bea \mathcal L_{\text{gf1}}&=& -2\left|D\cdot\vf^\mu-(1/2)D^\mu\vf'+ mB^\mu\right|^2,\label{rr16}\\
\mathcal L_{\text{gf2}}&=&-2\,|D\cdot B+(m/2)(\vf'- 3\phi)|^2.\eea{rr17}
One is thus left with
\beq \mathcal L=\vf_{\mu\nu}^{*}\left(\Box-m^2\right)\vf^{\mu\nu}-\tfrac{1}{2}\vf'^*\left(\Box-m^2\right)\vf'
+2B_\mu^*(\Box-m^2)B^\mu+\tfrac{3}{2}\phi^{*}(\Box-m^2)\phi-\tfrac{1}{4}F_{\mu\nu}^2+\mathcal L_{\text{int}}.\eeq{rr18}
Here $\mathcal L_{\text{int}}$ contains all interaction operators; they appear with canonical dimension ranging from 4
through 8, and at least one power of $e$. Parametrically in $e\ll1$, for any given operator dimensionality, the
$\mathcal{O}(e)$-terms are more dangerous than the others because they are weighted by the inverse of a lower mass
scale. On the other hand, at a given order in $e$, the higher the dimension, the more dangerous an operator is.
The most dangerous operators are therefore the $\mathcal{O}(e)$ dimension-8 ones:
\beq \mathcal L_8=\frac{e}{m^4}\,\partial_\mu F^{\mu\nu}[(i/2)\partial_\rho\phi^*\partial^\rho\partial_\nu\phi+\text{c.c.}]
\equiv \frac{e}{m^4}\,\partial_\mu F^{\mu\nu}J_\nu.\eeq{rr19}
While the field redefinition: $A_\mu\rightarrow A_\mu-(e/m^4)J_\mu$ removes this operator, it also yields another equally
bad term, namely, $(e/2m^4)^2(\partial_\mu J_\nu-\partial_\nu J_\mu)^2$. To improve the degree of divergence, one may add
some local functions of $\tilde{\vf}_{\mu\nu}$ that can cancel the $\mathcal O(e^2)$ term. Indeed the term
$\mathcal L_{\text{add}}=(e^2/4)(\tilde{\vf}^*_{\mu\rho}\tilde{\vf}^\rho_{~\nu}-\tilde{\vf}^*_{\nu\rho}\tilde{\vf}^\rho_{~\mu})^2$
achieves this feat. Thus the possible cutoff is pushed higher: the most dangerous operators now are weighted by $e^2/m^7$.
These dimension-11 operators could be eliminated, up to total derivatives, by adding local counter-terms. However, in a local
theory there is a cohomological obstruction to this possibility~\cite{pr2}. Attempts along these lines would therefore suggest
a cutoff scale $\mathcal O(m/e^{2/7})$, which cannot be improved further.

However, mere addition of a dipole term in the Lagrangian may leave us only with $\mathcal O(e/m^3)$-terms, as we will now see.
This is an improvement over field redefinition plus addition of local (counter-)terms. Indeed, the dipole term
\beq \mathcal L_{\text{dipole}}=ieF^{\mu \nu}\tilde{\vf}^*_{\mu\rho}\tilde{\vf}^\rho_{\;\;\nu},\eeq{dipole}
when added to the minimal Lagrangian~(\ref{rr7}), completely eliminates the dimension-8 operators~(\ref{rr19}). In the
resulting non-minimal theory, one does have $\mathcal{O}(e)$ dimension-7 operators:
\beq \mathcal L_7+\mathcal L_{7,\text{dipole}}=\frac{ie}{2m^3}\,F^{\mu\nu}\left(2\partial_{[\mu}
B_{\rho]}^*\partial^\rho\partial_\nu\phi-\partial_{[\mu}B_{\nu]}^*\Box\phi\right)+\text{c.c.},\eeq{rr30}
which contain pieces not proportional to any of the EoMs. Therefore, the degree of divergence cannot be improved
any further. In the scaling limit $m\rightarrow0$ and $e\rightarrow0$, such that $e/m^3=\text{constant}$, the non-minimal
Lagrangian reduces to:
\beq \mathcal L=\mathcal L_{\text{kin}}+\frac{i}{2\Lambda^3}\,F^{\mu\nu}\left(2\partial_{[\mu}B_{\rho]}^*\partial^\rho
\partial_\nu\phi-\partial_{[\mu}B_{\nu]}^*\Box\phi\right) + \text{c.c.},\eeq{rr31}
which has an intrinsic cutoff~\cite{pr1,pr2}:
\beq \Lambda= \frac{m}{e^{1/3}}.\eeq{rr32}

\subsection*{Arbitrary Spin}

One can follow the same steps for arbitrary spin to find that the largest cutoff of the local effective theory describing
a massive charged particle of $s\geq1$, coupled to EM is~\cite{pr2}:
\beq \Lambda=me^{-1/(2s-1)}\,.\eeq{m1}
This formula, valid in the limit $e\ll 1$, states that massive HS particles admit a local description even for energies
above their mass, when they must be treated as true dynamical DoFs in the effective action. This is a model-independent
upper bound on the cutoff that no theory can beat: requiring further consistencies, e.g. causal propagation, may only
result in a lower cutoff~\cite{AN2,PRS,pr2}. Conversely, other consistency requirements may even sharpen this bound.

The meaning of the cutoff upper bound~(\ref{m1}), as we have already mentioned, is only that some new physics must
happen below this scale. This new physics could result in a strong coupling unitarization, or in the existence of new
interacting DoFs lighter than the cutoff. In the first case, the theory becomes essentially non-local at a scale not
higher than $\Lambda$ as the spin-$s$ particle develops a form factor which is tantamount to a finite non-zero charge
radius. In the second case, if one integrates out the new light DoFs, the resulting action would necessarily contain
non-local terms already below the scale $\Lambda$. The very possibility non-local counter-terms invalidates the cohomological
argument of Ref.~\cite{pr2}. Because of this technical reason, lighter DoFs may be essential for a complete UV embedding of
the effective field theory.

\section{Interactions of Higher Spin Gauge Fields}\label{sec:six}

The no-go theorems for interacting HS fields, presented in Section~\ref{sec:no-go}, do not apply to massive particles.
The reason is trivial: gauge transformations are no longer a symmetry because of the mass term. However, we saw that
these fields have other problems when interactions are present. Some details of (consistent) interactions of massive
HS fields have been presented in Sections~\ref{sec:open} and~\ref{sec:five}.

The Aragone-Deser obstruction~\cite{ad} to consistent local interactions of massless HS fields is due
the appearance of the Weyl tensor in the gauge variation. In $D=3$, the Weyl tensor is vanishing, and so no such obstructions
exist. Because HS gauge fields in 3D carry no local DoFs (see Eqs.~(\ref{dofB0}) and~(\ref{dofF})), they are immune from
many consistency issues that may arise in higher dimensions. In fact, 3D HS gauge theory is a very rich and rapidly
expanding subject. An entire set of lecture notes by G.~Lucena~G\'omez on this topic$-$complementary to this
set$-$is also going to appear.

It is possible to have consistent HS gauge theories in AdS space, e.g., Vasiliev's HS gauge
theories~\cite{Vasiliev0,Vasiliev1}\footnote{See also Refs.~\cite{Ambient,Cubic_AdS0,Cubic_AdS} for discussions
of HS interactions in AdS space.}. Vasiliev's system is a set of classical non-linear gauge-invariant equations
for an infinite tower of HS gauge fields in AdS space. When expressed in terms of metric-like symmetric tensor
fields, the linearized equations take the standard Fronsdal form~(\ref{abba2}). These theories contain interaction
terms that do not stop at a finite number of derivatives, so that they are essentially non-local. But the non-locality
is under control in that higher-derivative terms are weighted by inverse powers of the cosmological constant $\L$,
which is at our disposal. However, this also means that these theories do not allow sensible flat limits. On the
other hand, the cosmological constant defines a mass parameter $\mathcal{O}(\sqrt{|\Lambda|})$: one expects no
operational difference between massless particles and massive particles with Compton wavelength larger than the AdS
radius. Another way of seeing how Vasiliev's theories avoid the no-go theorems is that in AdS space $S$-matrix is not
well-defined since asymptotic states do not exist. The formulation of these theories is rather technical;
interested readers may find Refs.~\cite{Misha,Carlo,GiombiReview} useful for a comprehensive overview.

Finally, in flat space another way to bypass the no-go theorems is to have higher-derivative non-minimal couplings for HS gauge
fields~\cite{p}. In fact, gravitational and EM multipole interactions show up, for example, as the $2-s-s$ and $1-s-s$
trilinear vertices constructed in~\cite{2-3-3} for bosonic fields. These cubic vertices are special cases of the general
form $s_1-s_2-s_3$, which involves massless fields of arbitrary spins. The light-cone formulation puts restrictions on
the number of derivatives in these vertices, and thereby provides a way of classifying them~\cite{Metsaev}. For
bosonic fields, there is one vertex for each value of the number of derivatives $p$, in the range
\beq s_1+s_2-s_3\leq p\leq s_1+s_2+s_3,\eeq{derivativeb} where $s_3$ is the smallest of the three spins. For a vertex
containing two fermions and a boson, the formula is the same modulo that one uses $s-\frac{1}{2}$ as ``spin'' for fermions.
For bosonic fields, the complete list of such vertices was given in~\cite{Cubic-general}. The authors in~\cite{Karapet}
have employed the Noether procedure to explicitly construct off-shell vertices. On the other hand, the tensionless limit of
String Theory gives rise to the same set of cubic vertices~\cite{Taronna}. In Ref.~\cite{Taronna} generating functions
for off-shell trilinear vertices for both bosonic and fermionic fields have been presented. One can also employ the
powerful machinery of the BRST-antifield method~\cite{BRST-BV} to systematically construct consistent interactions using
the language of cohomology. With the underlying assumptions of locality and Poincar\'e invariance, this approach is very
useful in general for obtaining gauge-invariant and manifestly Lorentz-invariant off-shell vertices for HS gauge
fields\footnote{See also Ref.~\cite{BBH} for general methodology and also~\cite{2-3-3,gravitons,N=1SUGRA,HLGR}
for some applications.}. In Section~\ref{sec:RS}, we will employ this method to construct a cubic coupling for a very
simple non-trivial system, while in Section~\ref{soc}, we explore the second-order consistency of this vertex.

\subsection{Consistent Couplings via BRST-Antifield Method}\label{sec:RS}

The BRST-antifield method~\cite{BRST-BV} uses gauge invariance as a consistency condition and by construction throws away
trivial interactions. Here we will employ this method to construct a parity-preserving off-shell non-abelian
$1-\tfrac{3}{2}-\tfrac{3}{2}$ vertex, closely following Ref.~\cite{HLGR}. The spin-$\tfrac{3}{2}$ system is simple enough
to allow an easy implementation of the BRST deformation scheme while capturing many non-trivial features that could serve
as guidelines for HS particles. Let us recall from Section~\ref{sec:no-go} that no-go theorems rule out minimal EM coupling
in flat space for $s=\tfrac{3}{2}$ and higher. To construct possible non-minimal couplings, we will go step by step for the
sake of a clear understanding of the procedure.

\subsection*{Step 0: Free Action}

The starting point is the free theory, which contains a photon $A_\mu$ and a massless spin-$\tfrac{3}{2}$ field $\psi_\mu$,
described by the action \beq S^{(0)}[A_\mu,\psi_\mu]=\int d^Dx\left[-\tfrac{1}{4}F_{\mu\nu}^2-i\bar{\psi}_\mu\gamma^{\mu\nu\rho}
\partial_\nu\psi_\rho\right],\eeq{rs1} which enjoys two abelian gauge invariances: one with a bosonic gauge parameter
$\lambda$, another with a fermionic gauge parameter $\varepsilon$: \beq \delta_\lambda A_\mu=\partial_\mu
\lambda,\qquad\delta_\varepsilon\psi_\mu=\partial_\mu\varepsilon.\eeq{rs2}

\subsection*{Step 1: Ghosts}

For the Grassmann-even gauge parameter $\lambda$, we introduce a Grassmann-odd bosonic ghost $C$. Corresponding to the
Grassmann-odd $\varepsilon$, we have a Grassmann-even fermionic ghost $\xi$. The set of fields now becomes
\beq \Phi^A=\{A_\mu, C, \psi_\mu, \xi\}.\eeq{rs3} In the algebra generated by the fields, we introduce a grading$-$the
pure ghost number ($pgh$)$-$which is non-zero only for the ghost fields $\{C, \xi\}$.

\subsection*{Step 2: Antifields}

For each field, we introduce an antifield with the same algebraic symmetries in its indices but opposite Grassmann parity.
The set of antifields is \beq \Phi^*_{A}=\{A^{*\mu}, C^*, \bar{\psi}^{*\mu},\bar{\xi}^*\}.\eeq{rs4}
In the algebra generated by the fields and antifields, we introduce another grading$-$the antighost number ($agh$)$-$which
is non-zero only for the antifields $\Phi^*_A$. Explicitly,
\beq agh(\Phi^*_A)=pgh(\Phi^A)+1,\qquad agh(\Phi^A)=0=pgh(\Phi^*_A).\eeq{agh}
The ghost number ($gh$) is another grading, defined as $gh=pgh-agh$.

\subsection*{Step 3: Antibracket}

On the space of fields and antifields we define an odd symplectic structure, called the antibracket, as
\beq (X,Y)\equiv\frac{\delta^RX}{\delta\Phi^A}\frac{\delta^LY}{\delta\Phi^*_A}-\frac{\delta^RX}{\delta\Phi^*_A}
\frac{\delta^LY}{\delta\Phi^A},\eeq{antibracket} which satisfies the graded Jacobi identity. This definition gives
$\left(\Phi^A,\Phi^*_B\right)=\delta^A_B$, which is real. Because a field and its antifield have opposite Grassmann parity,
it follows that if $\Phi^A$ is real, $\Phi^*_B$ must be purely imaginary, and vice versa.

\subsection*{Step 4: Free Master Action}

Now we construct the free master action $S_0$$-$an extension of the original gauge-invariant action~(\ref{rs1})$-$by
terms involving ghosts and antifields. Explicitly,
\beq S_0=\int d^Dx\left[-\tfrac{1}{4}F_{\mu\nu}^2-i\bar{\psi}_\mu\gamma^{\mu\nu\rho}\partial _\nu\psi_\rho+A^{*\mu}
\partial_\mu C - (\bar{\psi}^{*\mu}\partial_\mu\xi-\partial_\mu\bar{\xi}\psi^{*\mu})\right].\eeq{rs5} Note that the
antifields appear as sources for the ``gauge'' variations, with gauge parameters replaced by corresponding ghosts.
It is easy to see that~(\ref{rs5}) solves the master equation $(S_0,S_0)=0$.

\subsection*{Step 5: BRST Differential}

The master action $S_0$ is the generator of the BRST differential $s$ for the free theory$-$defined as
\beq sX\equiv(S_0,X).\eeq{brst1} Note that $S_0$ is BRST-closed, as a simple consequence of the master equation.
From the properties of the antibracket, we also find that $s$ is nilpotent:
\beq s^2=0.\eeq{brst2}
Then, the master action $S_0$ belongs to the cohomology of $s$ in the space of local functionals of the fields,
antifields, and their finite number of derivatives.

The various gradings are important as $s$ decomposes into the sum of the Koszul-Tate differential,
$\Delta$, and the longitudinal derivative along the gauge orbits, $\Gamma$: \beq s=\Delta+\Gamma. \eeq{brst3}
$\Delta$ implements the EoMs by acting only on the antifields, and in so doing it decreases the antighost
number by one unit while keeping unchanged the pure ghost number. On the other hand, $\Gamma$ acts only on the
original fields and produces the gauge transformations. It increases the pure ghost number by one unit
without changing the antighost number. Thus all three $\Delta$, $\Gamma$ and $s$ increase the ghost number
by one unit, $gh(\Delta)=gh(\Gamma)=gh(s)=1$. It is important to note that $\Delta$ and $\Gamma$ are nilpotent
and anticommuting: \beq \Gamma^2=\Delta^2=0,\qquad\Gamma\Delta+\Delta\Gamma=0.\eeq{brst4}

The different gradings and Grassmann parity of the various fields and antifields, along with the action of
$\Gamma$ and $\Delta$ on them, are given in the following Table.

\begin{table}[ht]
\caption{Properties of the Various Fields \& Antifields}
\vspace{6pt}
\centering
\begin{tabular}{c c c c c c c}
\hline\hline
$Z$ &$\Gamma(Z)$~~~&~~~$\Delta(Z)$~~~&$pgh(Z)$ &$agh(Z)$ &$gh(Z)$ &$\epsilon(Z)$\\ [0.5ex]
\hline
$A_\mu$ & $\partial_\mu C$ & 0 & 0 & 0 & 0 & 0\\
$C$ & 0 & 0 & 1 & 0 & 1 & 1\\
$A^{*\mu}$ & 0 & $-\partial_\nu F^{\mu\nu}$ & 0 & 1 & $-1$ & 1\\
$C^*$ & 0 & $-\partial_\mu A^{*\mu}$ & 0 & 2 & $-2$ & 0\\ \hline
$\psi_\mu$ & $\partial_\mu\xi$ & 0 & 0 & 0 & 0 & 1\\
$\xi$ & 0 & 0 & 1 & 0 & 1 & 0\\
$\bar{\psi}^{*\mu}$ & 0 & $-i\de_\alpha\bar\psi_\beta\gamma^{\alpha\beta\mu}$ & 0 & 1 & $-1$ & 0\\
$\bar{\xi}^*$ & 0 & $\partial_\mu\bar{\psi}^{*\mu}$ & 0 & 2 & $-2$ & 1\\
\hline\hline
\end{tabular}
\end{table}
\vspace{6pt}

\subsection*{An Aside: BRST Deformation Scheme}

The solution of the master equation contains compactly all information pertaining to the consistency of the gauge transformations.
One can reformulate the problem of introducing consistent interactions in a gauge theory as that of deforming the master action.
If $S$ is the solution of the deformed master equation, then $(S,S)=0$. This must be a deformation of the solution $S_0$
of the master equation of the free gauge theory, in the deformation parameter $g$:
\beq S=S_0+gS_1+g^2S_2+\mathcal O(g^3).\eeq{brst5} The master equation for $S$ splits, up to $\mathcal O(g^2)$,
into the following set \bea &(S_0,S_0)=0,&\label{brst6.1}\\&(S_0,S_1)=0,&\label{brst6.2}\\&(S_1,S_1)=-2(S_0,S_2).&\eea{brst6.3}
Eq.~(\ref{brst6.1}) is fulfilled by assumption, and Eq.~(\ref{brst6.2}) means $S_1$ is BRST-closed:
\beq s S_1=0.\eeq{brst1st} If the first-order local deformations are given by $S_1=\int a$, where $a$ is a top-form of ghost
number 0, then the following cocycle condition follows \beq sa+db=0,\eeq{cocycle0} which says that
non-trivial deformations belong to $H^0(s|d)$$-$the cohomology of the zeroth-order BRST differential $s$, modulo total
derivative $d$, at ghost number 0. Now, one can make an antighost-number expansion of the local form $a$; this expansion stops
at $agh=2$~\cite{2-3-3,BBH,gravitons}:
\beq a=a_0+a_1+a_2, \qquad agh(a_i)=i=pgh(a_i).\eeq{brst7}
The significance of the various terms is as follows. $a_0$ is the deformation of the Lagrangian, while $a_1$
and $a_2$ encode information about the deformations of the gauge transformations and the gauge algebra
respectively~\cite{BRST-BV}. Thus, if $a_2$ is non-trivial, the algebra of the gauge transformations is deformed and
becomes non-abelian. In what follows we are interested only in gauge-algebra-deforming/non-abelian vertices.
Given the expansion~(\ref{brst7}) and the decomposition~(\ref{brst3}), the cocycle condition~(\ref{cocycle0}) gives
the following consistency cascade for allowed deformations:
\bea &\Gamma a_2=0,\label{cocycle1}&\\
&\Delta a_2+\Gamma a_1+db_1=0,&\label{cocycle2}\\
&\Delta a_1+\Gamma a_0+db_0=0,&\eea{cocycle3}
where $agh(b_i)=i,~pgh(b_i)=i+1$, and $a_2$ has been chosen to be $\Gamma\text{-closed}$, instead of $\Gamma\text{-closed}$
modulo $d$ (this is always possible~\cite{BBH}).

\subsection*{Step 6: Cohomology of $\Gamma$}

The cohomology of $\Gamma$ contains gauge-invariant functions of the fields and antifields.
For the spin-$\tfrac{3}{2}$ field, it is isomorphic to the space of functions of
\begin{itemize}
 \item The undifferentiated ghosts $\{C, \xi\}$,
 \item The antifields $\{A^{*\mu}, C^*, \bar{\psi}^{*\mu},\bar{\xi}^*\}$ and their derivatives,
 \item The curvatures $\{F_{\mu\nu}, \de_{[\mu}\psi_{\nu]}\}$ and their derivatives.
\end{itemize}

\subsection*{Step 7: Solution of the Consistency Cascade}

Requiring that $a_2$ be a parity-even Lorentz scalar, the most general possibility for it is
\beq a_2=-g_0C\left(\bar{\xi}^*\xi+\bar{\xi}\xi^*\right)-g_1C^*\bar{\xi}\xi,\eeq{rs6}
The first piece with coefficient $g_0$ potentially gives rise to minimal coupling, while the second could produce
dipole interaction. This can be understood by a derivative counting down the cascade~(\ref{cocycle1})-(\ref{cocycle3})
with the knowledge of the action of $\Gamma$ and $\Delta$ on the fields and antifields. It is then easy to see that
the respective $a_0$'s would contain no derivative and one derivative respectively.

It turns out that each of the terms in $a_2$ can be consistently lifted to an $a_1$ by solving Eq.~(\ref{cocycle2}):
\beq a_1=g_0\left[\bar{\psi}^{*\mu}(\psi_\mu C+\xi A_\mu)+\text{h.c.}\right]+g_1A^{*\mu}(\bar{\psi}_\mu\xi-\bar{\xi}\psi_\mu)
+\tilde{a}_1,\qquad \Gamma\tilde{a}_1=0,\eeq{rs8} where the ambiguity, $\tilde{a}_1$, belongs to the cohomology of $\Gamma$.
Now $\Delta a_1$ is by construction $\Gamma$-closed modulo $d$, but condition~(\ref{cocycle3}) requires it to be
$\Gamma$-exact modulo $d$. This can happen only when in $\Delta a_1$ the potential $\Gamma$-non-trivial pieces coming from
the unambiguous piece and the ambiguity cancel each other. It is easy to see that such a cancelation is impossible for
the first term in $a_1$ simply because $\tilde{a}_1$ contains too many derivatives. This rules out minimal coupling of a
massless spin-$\tfrac{3}{2}$ to EM, in accordance with what is known~\cite{p,Metsaev}.

With $g_0=0$, one indeed can satisfy Eq.~(\ref{cocycle3}) with the choice
\beq \tilde{a_1}=ig_1\left[\bar{\psi}^{*\mu}\gamma^\nu F_{\mu\nu}-\tfrac{1}{2(D-2)}\bar{\displaystyle\not{\!\!\psi}}^*
\displaystyle{\not{\!F}}\right]\xi+\text{h.c.},\eeq{rs13}
which is of course in the cohomology of $\Gamma$. Thus we have a consistent Lagrangian deformation
\beq a_0=g_1\bar{\psi}_\mu F^{+\mu\nu}\psi_\nu=g_1\bar{\psi}_\mu\left(F^{\mu\nu}+\tfrac{1}{2}\gamma^{\mu\nu\r\s}F_{\r\s}\right)
\psi_\nu,\eeq{rs15} which is the desired non-abelian vertex.

\subsection{Second Order Consistency}\label{soc}

We recall that consistent second-order deformation requires that $(S_1,S_1)$ be $s$-exact: \beq (S_1,S_1)=-2sS_2=
-2\Delta S_2 -2\Gamma S_2.\eeq{2nd1}
We can prove by contradiction that there is obstruction for our cubic non-abelian vertex to the quartic order.
If Eq.~(\ref{2nd1}) holds, then one can show that the $\Gamma$-variation of the most general form
of this antibracket evaluated at zero antifields is $\Delta$-exact~\cite{HLGR}:
\beq \Gamma\left[(S_1,S_1)\right]_{\Phi^*_A=0}=\Gamma\Delta M=-\Delta\left(\Gamma M\right).\eeq{2nd3}
It is relatively easier to compute the left hand side of~(\ref{2nd3}) for our non-abelian vertex.
Explicit computation gives \bea \left[\left(S_1,S_1\right)\right]_{\Phi^*_A=0}&=&\int d^Dx\left\{4(\bar{\psi}_\mu
\xi-\bar{\xi}\psi_\mu)\,\partial_\nu\left(\bar{\psi}^{[\mu}\psi^{\nu]}+\tfrac{1}{2}\bar{\psi}_\alpha\gamma^{\mu\nu\alpha\beta}
\psi_\beta\right)\right\}\nonumber\\&&+\int d^Dx\left\{i\bar{\psi}_\mu F^{+\mu\nu}\left[2\gamma^\rho F_{\nu\rho}-\tfrac{1}
{D-2}\gamma_\nu\displaystyle{\not{\!F}}\right]\xi+\text{h.c.}\right\}.\eea{2nd8} If the vertex is unobstructed so that
Eq.~(\ref{2nd3}) holds, then the $\Gamma$-variation of each of these terms should separately be $\Delta$-exact. By
considering, for example, the fermion bilinears appearing in the second line of Eq.~(\ref{2nd8}) it is easy to see that
their $\Gamma$-variation is not $\Delta$-exact. The conclusion is that the non-abelian $1-\tfrac{3}{2}-\tfrac{3}{2}$
vertex gets obstructed beyond the cubic order.

Notice that the vertex $\bar{\psi}_\mu F^{+\mu\nu}\psi_\nu$ is just the Pauli term appearing in $\mathcal N=2$
SUGRA~\cite{SUGRA1}. This theory, however, contains additional DoFs$-$a graviton$-$beside a complex massless
spin $\tfrac{3}{2}$ and a $U(1)$ field. It is this new dynamical field that removes obstructions to the vertex while keeping
locality intact. One may decouple gravity by taking $M_\text{P}\rightarrow\infty$; then the Pauli term vanishes because the
dimensionful coupling constant goes like $1/M_\text{P}$~\cite{SUGRA1}. On the other hand, one could integrate out the massless
graviton to obtain a system of spin-$\tfrac{3}{2}$ and spin-1 fields only. The resulting theory does contain the Pauli term,
but it is necessarily non-local. Thus, higher-order consistency of the non-abelian vertex requires that one either forgo
locality or add a new dynamical field (graviton). For higher spin values, higher-order consistency may call for an infinite
tower of HS gauge fields as well as non-locality just like in Vasiliev's theory~\cite{Vasiliev0,Vasiliev1} or the tensionless
limit of String Theory.

\section*{Acknowledgments}

We would like to thank A.~Campoleoni for useful discussions. RR is a Postdoctoral Fellow of the Fonds de la Recherche
Scientifique-FNRS. His work is partially supported by IISN-Belgium (conventions 4.4511.06 and 4.4514.08) and by the
``Communaut\'e Fran\c{c}aise de Belgique'' through the ARC program and by the ERC Advanced Grant ``SyDuGraM.''


\begin{thebibliography}{99}

\bibitem{w64}
  S.~Weinberg,
  ``Photons And Gravitons In S Matrix Theory: Derivation Of Charge Conservation
  And Equality Of Gravitational And Inertial Mass,''
  Phys.\ Rev.\  {\bf 135}, B1049 (1964).

\bibitem{gpv}
  M.~T.~Grisaru and H.~N.~Pendleton,
  ``Soft Spin 3/2 Fermions Require Gravity And Supersymmetry,''
  Phys.\ Lett.\  B {\bf 67}, 323 (1977);
  M.~T.~Grisaru, H.~N.~Pendleton and P.~van Nieuwenhuizen,
  Phys.\ Rev.\  D {\bf 15}, 996 (1977).

\bibitem{ad}
  C.~Aragone and S.~Deser,
  ``Consistency Problems Of Hypergravity,''
  Phys.\ Lett.\  B {\bf 86}, 161 (1979).

\bibitem{ww}
  S.~Weinberg and E.~Witten,
  ``Limits On Massless Particles,''
  Phys.\ Lett.\  B {\bf 96}, 59 (1980).

\bibitem{p}
  M.~Porrati,
  ``Universal Limits on Massless High-Spin Particles,''
  Phys. \ Rev. \ D {\bf 78}, 065016 (2008)
  [arXiv:0804.4672 [hep-th]].

\bibitem{Gross}
  D.~J.~Gross and P.~F.~Mende,
  ``The High-Energy Behavior of String Scattering Amplitudes,''
  Phys.\ Lett.\ B {\bf 197}, 129 (1987),
  ``String Theory Beyond the Planck Scale,''
  Nucl.\ Phys.\ B {\bf 303}, 407 (1988);
  D.~J.~Gross,
  ``High-Energy Symmetries of String Theory,''
  Phys.\ Rev.\ Lett.\  {\bf 60}, 1229 (1988).

\bibitem{Amati}
  D.~Amati, M.~Ciafaloni and G.~Veneziano,
  ``Superstring Collisions at Planckian Energies,''
  Phys.\ Lett.\ B {\bf 197}, 81 (1987),
  ``Classical and Quantum Gravity Effects from Planckian Energy Superstring Collisions,''
  Int.\ J.\ Mod.\ Phys.\ A {\bf 3}, 1615 (1988),
  ``Can Space-Time Be Probed Below the String Size?,''
  Phys.\ Lett.\ B {\bf 216}, 41 (1989).

\bibitem{Vasiliev0}
  E.~S.~Fradkin and M.~A.~Vasiliev,
  ``On The Gravitational Interaction Of Massless Higher Spin Fields,''
  Phys.\ Lett.\ B {\bf 189}, 89 (1987),
  ``Cubic Interaction In Extended Theories Of Massless Higher Spin Fields,''
  Nucl.\ Phys.\ B {\bf 291}, 141 (1987).

\bibitem{Sagnotti11}
  A.~Sagnotti,
  ``Notes on Strings and Higher Spins,''
  J.\ Phys.\ A {\bf 46}, 214006 (2013)
  [arXiv:1112.4285 [hep-th]].

\bibitem{Vasiliev1}
  M.~A.~Vasiliev,
  ``Consistent equation for interacting gauge fields of all spins in
  (3+1)-dimensions,''
  Phys.\ Lett.\ B {\bf 243}, 378 (1990),
  ``Properties of equations of motion of interacting gauge fields of all spins
  in (3+1)-dimensions,''
  Class.\ Quant.\ Grav.\  {\bf 8}, 1387 (1991),
  ``More On Equations Of Motion For Interacting Massless Fields Of All Spins In (3+1)-Dimensions,''
  Phys.\ Lett.\ B {\bf 285}, 225 (1992),
  ``Nonlinear equations for symmetric massless higher spin fields in (A)dS(d),''
  Phys.\ Lett.\  B {\bf 567}, 139 (2003)
  [arXiv:hep-th/0304049].

\bibitem{KP}
  I.~R.~Klebanov and A.~M.~Polyakov,
  ``AdS dual of the critical O(N) vector model,''
  Phys.\ Lett.\ B {\bf 550}, 213 (2002)
  [hep-th/0210114].
  E.~Sezgin and P.~Sundell,
  ``Holography in 4D (super) higher spin theories and a test via cubic scalar couplings,''
  JHEP {\bf 0507}, 044 (2005)
  [hep-th/0305040].

\bibitem{GiombiYin}
  S.~Giombi and X.~Yin,
  ``Higher Spin Gauge Theory and Holography: The Three-Point Functions,''
  JHEP {\bf 1009}, 115 (2010)
  [arXiv:0912.3462 [hep-th]];
  ``Higher Spins in AdS and Twistorial Holography,''
  JHEP {\bf 1104}, 086 (2011)
  [arXiv:1004.3736 [hep-th]].

\bibitem{vz}
  G.~Velo and D.~Zwanziger,
  ``Propagation And Quantization Of Rarita-Schwinger Waves In An External Electromagnetic Potential,''
  Phys.\ Rev.\  {\bf 186}, 1337 (1969),
  ``Noncausality and other defects of interaction lagrangians for particles with spin one and higher,''
  Phys.\ Rev.\  {\bf 188}, 2218 (1969);
  G.~Velo,
  ``Anomalous behaviour of a massive spin two charged particle in an external electromagnetic field,''
  Nucl.\ Phys.\  B {\bf 43}, 389 (1972).

\bibitem{dirac}
  P.~A.~M.~Dirac,
  ``Relativistic wave equations,''
  Proc.\ Roy.\ Soc.\ Lond.\  {\bf 155A}, 447 (1936).

\bibitem{pf}
  M.~Fierz and W.~Pauli,
  ``On relativistic wave equations for particles of arbitrary spin in an electromagnetic field,''
  Proc.\ Roy.\ Soc.\ Lond.\  A {\bf 173}, 211 (1939).

\bibitem{wig1}
  E.~P.~Wigner,
  ``On Unitary Representations Of The Inhomogeneous Lorentz Group,''
  Annals Math.\  {\bf 40}, 149 (1939)
  [Nucl.\ Phys.\ Proc.\ Suppl.\  {\bf 6}, 9 (1989)].

\bibitem{wig2}
  V.~Bargmann and E.~P.~Wigner,
  ``Group Theoretical Discussion Of Relativistic Wave Equations,''
  Proc.\ Nat.\ Acad.\ Sci.\  {\bf 34}, 211 (1948).

\bibitem{frons}
  C.~Fronsdal,
  Nuovo Cim.\ Suppl.\ {\bf 9}, 416 (1958).

\bibitem{chang}
  S.~J.~Chang,
  ``Lagrange Formulation for Systems with Higher Spin,''
  Phys.\ Rev.\  {\bf 161}, 1308 (1967).

\bibitem{sh}
  L.~P.~S.~Singh and C.~R.~Hagen,
  ``Lagrangian formulation for arbitrary spin. 1. The boson case,''
  Phys.\ Rev.\  D {\bf 9}, 898 (1974),
  ``Lagrangian formulation for arbitrary spin. 2. The fermion case,''
  Phys.\ Rev.\  D {\bf 9}, 910 (1974).

\bibitem{ff}
  C.~Fronsdal,
  ``Massless Fields With Integer Spin,''
  Phys.\ Rev.\  D {\bf 18}, 3624 (1978);
  J.~Fang and C.~Fronsdal,
  ``Massless Fields With Half Integral Spin,''
  Phys.\ Rev.\  D {\bf 18}, 3630 (1978).

\bibitem{deW}
  B.~de Wit and D.~Z.~Freedman,
  ``Systematics Of Higher Spin Gauge Fields,''
  Phys.\ Rev.\ D {\bf 21}, 358 (1980).

\bibitem{Dario}
  D.~Francia and A.~Sagnotti,
  ``Free geometric equations for higher spins,''
  Phys.\ Lett.\ B {\bf 543}, 303 (2002)  [hep-th/0207002],
  ``On the geometry of higher spin gauge fields,''
  Class.\ Quant.\ Grav.\  {\bf 20}, S473 (2003)  [hep-th/0212185].

\bibitem{ady}
  C.~Aragone, S.~Deser and Z.~Yang,
  ``MASSIVE HIGHER SPIN FROM DIMENSIONAL REDUCTION OF GAUGE FIELDS,''
  Annals Phys.\  {\bf 179}, 76 (1987);
  S.~D.~Rindani and M.~Sivakumar,
  ``Gauge - Invariant Description Of Massive Higher - Spin Particles By Dimensional Reduction,''
  Phys.\ Rev.\  D {\bf 32}, 3238 (1985);
  S.~D.~Rindani, D.~Sahdev and M.~Sivakumar,
  ``Dimensional reduction of symmetric higher spin actions. 1. Bosons,''
  Mod.\ Phys.\ Lett.\  A {\bf 4}, 265 (1989),
  ``Dimensional Reduction Of Symmetric Higher Spin Actions. 2: Fermions,''
  Mod.\ Phys.\ Lett.\  A {\bf 4}, 275 (1989).

\bibitem{dw2}
  S.~Deser and A.~Waldron,
  ``Stress and strain: T(mu nu) of higher spin gauge fields,''
  arXiv:hep-th/0403059.

\bibitem{deser}
  S.~Deser and A.~Waldron,
  ``Inconsistencies of massive charged gravitating higher spins,''
  Nucl.\ Phys.\ B {\bf 631}, 369 (2002)
  [hep-th/0112182].

\bibitem{fed}
  P. Federbush,
  ``Minimal Electromagnetic Coupling for Spin Two Particles,''
  Nuovo Cimento {\bf 19}, 572 (1961).

\bibitem{schwinger}
  J.~S.~Schwinger,
  ``On gauge invariance and vacuum polarization,''
  Phys.\ Rev.\  {\bf 82}, 664  (1951);
  C.~Bachas and M.~Porrati,
  ``Pair creation of open strings in an electric field,''
  Phys.\ Lett.\  B {\bf 296}, 77 (1992)
  [arXiv:hep-th/9209032].

\bibitem{nielsen}
  N.~K.~Nielsen and P.~Olesen,
  ``An Unstable Yang-Mills Field Mode,''
  Nucl.\ Phys.\  B {\bf 144}, 376 (1978).

\bibitem{js}
  K.~Johnson and E.~C.~G.~Sudarshan,
  ``Inconsistency of the local field theory of charged spin 3/2 particles,''
  Annals Phys.\  {\bf 13}, 126 (1961).

\bibitem{same}
  J.~D.~Jenkins,
  ``Constraints, Causality And Lorentz Invariance,''
  J.\ Phys.\ A  {\bf 7}, 1129 (1974);
  M.~Kobayashi and Y.~Takahashi,
  ``Origin Of The Gribov Ambiguity,''
  Phys.\ Lett.\  B {\bf 78}, 241 (1978),
  ``THE RARITA-SCHWINGER PARADOXES,''
  J.\ Phys.\ A  {\bf 20}, 6581 (1987).

\bibitem{pr1}
  M.~Porrati and R.~Rahman,
  ``Intrinsic Cutoff and Acausality for Massive Spin 2 Fields Coupled to Electromagnetism,''
  Nucl.\ Phys.\  B {\bf 801}, 174 (2008)
  [arXiv:0801.2581 [hep-th]],
  ``Electromagnetically Interacting Massive Spin-2 Field: Intrinsic Cutoff and Pathologies in External Fields,''
  arXiv:0809.2807 [hep-th].

\bibitem{kob}
  M.~Kobayashi and A.~Shamaly,
  ``Minimal Electromagnetic Coupling For Massive Spin-2 Fields,''
  Phys.\ Rev.\  D {\bf 17}, 2179 (1978),
  ``The Tenth Constraint In The Minimally Coupled Spin-2 Wave Equations,''
  Prog.\ Theor.\ Phys.\  {\bf 61}, 656 (1979).

\bibitem{d3}
  S.~Deser, V.~Pascalutsa and A.~Waldron,
  ``Massive spin 3/2 electrodynamics,''
  Phys.\ Rev.\  D {\bf 62}, 105031 (2000)
  [arXiv:hep-th/0003011].

\bibitem{sham}
  A.~Shamaly and A.~Z.~Capri,
  ``Propagation Of Interacting Fields,''
  Annals Phys.\  {\bf 74}, 503 (1972);
  M.~Hortacsu,
  ``Demonstration of noncausality for the rarita-schwinger equation,''
  Phys.\ Rev.\ D {\bf 9}, 928 (1974).


\bibitem{PR2}
  M.~Porrati, R.~Rahman,
  ``Causal Propagation of a Charged Spin 3/2 Field in an External Electromagnetic Background,''
  Phys.\ Rev.\  {\bf D80}, 025009 (2009).
  [arXiv:0906.1432 [hep-th]].

\bibitem{SUGRA1}
  S.~Ferrara and P.~van Nieuwenhuizen,
  ``Consistent Supergravity With Complex Spin 3/2 Gauge Fields,''
  Phys.\ Rev.\ Lett.\  {\bf 37}, 1669 (1976);
  D.~Z.~Freedman, A.~K.~Das,
  ``Gauge Internal Symmetry in Extended Supergravity,''
  Nucl.\ Phys.\  {\bf B120}, 221 (1977).

\bibitem{SUGRA2}
  J.~Scherk and J.~H.~Schwarz,
  ``Spontaneous Breaking Of Supersymmetry Through Dimensional Reduction,''
  Phys.\ Lett.\  B {\bf 82}, 60 (1979),
  ``How To Get Masses From Extra Dimensions,''
  Nucl.\ Phys.\  B {\bf 153}, 61 (1979);
  B.~de Wit, P.~G.~Lauwers and A.~Van Proeyen,
  ``Lagrangians Of N=2 Supergravity - Matter Systems,''
  Nucl.\ Phys.\  B {\bf 255}, 569 (1985).

\bibitem{DZ}
  S.~Deser and B.~Zumino,
  ``Broken Supersymmetry And Supergravity,''
  Phys.\ Rev.\ Lett.\  {\bf 38}, 1433 (1977).

\bibitem{Abouelsaood}
E.~S.~Fradkin and A.~A.~Tseytlin,
  ``Nonlinear Electrodynamics From Quantized Strings,''
  Phys.\ Lett.\  B {\bf 163}, 123 (1985);
  A.~Abouelsaood, C.~G.~.~Callan, C.~R.~Nappi and S.~A.~Yost,
  ``Open Strings In Background Gauge Fields,''
  Nucl.\ Phys.\  B {\bf 280}, 599 (1987).

\bibitem{AN2}
  P.~C.~Argyres and C.~R.~Nappi,
  ``MASSIVE SPIN-2 BOSONIC STRING STATES IN AN ELECTROMAGNETIC BACKGROUND,''
  Phys.\ Lett.\  B {\bf 224}, 89 (1989).

\bibitem{AN1}
  P.~C.~Argyres and C.~R.~Nappi,
  ``Spin 1 Effective Actions From Open Strings,''
  Nucl.\ Phys.\  B {\bf 330}, 151 (1990).

\bibitem{PRS}
  M.~Porrati, R.~Rahman and A.~Sagnotti,
  ``String Theory and The Velo-Zwanziger Problem,''
  Nucl.\ Phys.\ B {\bf 846}, 250 (2011)
  [arXiv:1011.6411 [hep-th]].

\bibitem{string3}
  S.~M.~Klishevich,
  ``Electromagnetic interaction of massive spin-3 state from string theory,''
  Int.\ J.\ Mod.\ Phys.\  A {\bf 15}, 395 (2000)
  [arXiv:hep-th/9805174].

\bibitem{2plus0}
  M.~Porrati and R.~Rahman,
  ``Notes on a Cure for Higher-Spin Acausality,''
  Phys.\ Rev.\ D {\bf 84}, 045013 (2011)
  [arXiv:1103.6027 [hep-th]].

\bibitem{Ferrara}
  S.~Ferrara, M.~Porrati and V.~L.~Telegdi,
  ``g = 2 as the natural value of the tree level gyromagnetic ratio of elementary particles,''
  Phys.\ Rev.\ D {\bf 46}, 3529 (1992).


\bibitem{Nima}
  N.~Arkani-Hamed, H.~Georgi and M.~D.~Schwartz,
  ``Effective field theory for massive gravitons and gravity in theory space,''
  Annals Phys.\  {\bf 305}, 96 (2003)
  [hep-th/0210184];

\bibitem{Claudia}
  C.~de Rham, G.~Gabadadze and A.~J.~Tolley,
  ``Ghost free Massive Gravity in the St\'uckelberg language,''
  Phys.\ Lett.\ B {\bf 711}, 190 (2012)
  [arXiv:1107.3820 [hep-th]],
  ``Helicity Decomposition of Ghost-free Massive Gravity,''
  JHEP {\bf 1111}, 093 (2011)
  [arXiv:1108.4521 [hep-th]].

\bibitem{pr2}
  M.~Porrati and R.~Rahman,
  ``A Model Independent Ultraviolet Cutoff for Theories with Charged Massive Higher Spin Fields,''
  Nucl.\ Phys.\ B {\bf 814}, 370 (2009)
  [arXiv:0812.4254 [hep-th]].



\bibitem{Ambient}
  R.~R.~Metsaev,
  Phys.\ Lett.\ B {\bf 354}, 78 (1995);
  Phys.\ Lett.\ B {\bf 419}, 49 (1998)  [hep-th/9802097].
  A.~Fotopoulos, K.~L.~Panigrahi and M.~Tsulaia,
  Phys.\ Rev.\ D {\bf 74}, 085029 (2006)  [hep-th/0607248].

\bibitem{Cubic_AdS0}
  K.~Alkalaev,
  ``FV-type action for $AdS_5$ mixed-symmetry fields,''
  JHEP {\bf 1103}, 031 (2011)  [arXiv:1011.6109 [hep-th]];
  N.~Boulanger and E.~D.~Skvortsov,
  ``Higher-spin algebras and cubic interactions for simple mixed-symmetry fields in AdS spacetime,''
  JHEP {\bf 1109}, 063 (2011)  [arXiv:1107.5028 [hep-th]];
  N.~Boulanger, E.~D.~Skvortsov and Y.~.M.~Zinoviev,
  ``Gravitational cubic interactions for a simple mixed-symmetry gauge field in AdS and flat backgrounds,''
  J.\ Phys.\ A A {\bf 44}, 415403 (2011)  [arXiv:1107.1872 [hep-th]];
  M.~A.~Vasiliev,
  ``Cubic Vertices for Symmetric Higher-Spin Gauge Fields in $(A)dS_d$,''
  arXiv:1108.5921 [hep-th].

\bibitem{Cubic_AdS}
  E.~Joung and M.~Taronna,
  ``Cubic interactions of massless higher spins in (A)dS: metric-like approach,''
  Nucl.\ Phys.\ B {\bf 861}, 145 (2012)  [arXiv:1110.5918 [hep-th]];
  E.~Joung, L.~Lopez and M.~Taronna,
  ``On the cubic interactions of massive and partially-massless higher spins in (A)dS,''
  JHEP {\bf 1207}, 041 (2012)
  [arXiv:1203.6578 [hep-th]].

\bibitem{Misha}
  M.~A.~Vasiliev,
  ``Higher spin gauge theories in four-dimensions, three-dimensions, and two-dimensions,''
  Int.\ J.\ Mod.\ Phys.\ D {\bf 5}, 763 (1996)
  [hep-th/9611024],
  ``Higher spin gauge theories: Star product and AdS space,''
  In *Shifman, M.A. (ed.): The many faces of the superworld* 533-610
  [hep-th/9910096].

\bibitem{Carlo}
  X.~Bekaert, S.~Cnockaert, C.~Iazeolla and M.~A.~Vasiliev,
  ``Nonlinear higher spin theories in various dimensions,''
  arXiv:hep-th/0503128;
  C.~Iazeolla,
  ``On the Algebraic Structure of Higher-Spin Field Equations and New Exact Solutions,''
  arXiv:0807.0406 [hep-th].


\bibitem{GiombiReview}
  S.~Giombi and X.~Yin,
  ``The Higher Spin/Vector Model Duality,''
  J.\ Phys.\ A {\bf 46}, 214003 (2013)
  [arXiv:1208.4036 [hep-th]].

\bibitem{2-3-3}
  N.~Boulanger and S.~Leclercq,
  ``Consistent couplings between spin-2 and spin-3 massless fields,''
  JHEP {\bf 0611}, 034 (2006)  [hep-th/0609221];
  N.~Boulanger, S.~Leclercq and P.~Sundell,
  ``On The Uniqueness of Minimal Coupling in Higher-Spin Gauge Theory,''
  JHEP {\bf 0808}, 056 (2008)  [arXiv:0805.2764 [hep-th]].

\bibitem{Metsaev}
  R.~R.~Metsaev,
  ``Cubic interaction vertices of massive and massless higher spin fields,''
  Nucl.\ Phys.\ B {\bf 759}, 147 (2006)  [hep-th/0512342],
  ``Cubic interaction vertices for fermionic and bosonic arbitrary spin fields,''
  Nucl.\ Phys.\ B {\bf 859}, 13 (2012)  [arXiv:0712.3526 [hep-th]].

\bibitem{Cubic-general}
  A.~K.~H.~Bengtsson, I.~Bengtsson and N.~Linden,
  ``Interacting Higher Spin Gauge Fields On The Light Front,''
  Class.\ Quant.\ Grav.\  {\bf 4}, 1333 (1987).

\bibitem{Karapet}
  R.~Manvelyan, K.~Mkrtchyan and W.~Ruhl,
  ``General trilinear interaction for arbitrary even higher spin gauge fields,''
  Nucl.\ Phys.\ B {\bf 836}, 204 (2010)  [arXiv:1003.2877 [hep-th]],
  ``A Generating function for the cubic interactions of higher spin fields,''
  Phys.\ Lett.\ B {\bf 696}, 410 (2011)  [arXiv:1009.1054 [hep-th]].

\bibitem{Taronna}
  A.~Sagnotti and M.~Taronna,
  `String Lessons for Higher-Spin Interactions,''
  Nucl.\ Phys.\ B {\bf 842}, 299 (2011)  [arXiv:1006.5242 [hep-th]].

\bibitem{BRST-BV}
  G.~Barnich and M.~Henneaux,
  ``Consistent couplings between fields with a gauge freedom and deformations of the master equation,''
  Phys.\ Lett.\ B {\bf 311}, 123 (1993)  [hep-th/9304057].
  M.~Henneaux,
  ``Consistent interactions between gauge fields: The Cohomological approach,''
  Contemp.\ Math.\  {\bf 219}, 93 (1998)  [hep-th/9712226].

\bibitem{BBH}
  G.~Barnich, F.~Brandt and M.~Henneaux,
  ``Local BRST cohomology in the antifield formalism. 1. General theorems,''
  Commun.\ Math.\ Phys.\  {\bf 174}, 57 (1995)  [hep-th/9405109];
  ``Local BRST cohomology in the antifield formalism. II. Application to Yang-Mills theory,''
  Commun.\ Math.\ Phys.\  {\bf 174}, 93 (1995)  [hep-th/9405194];
  ``Conserved currents and gauge invariance in Yang-Mills theory,''
  Phys.\ Lett.\ B {\bf 346}, 81 (1995)  [hep-th/9411202];
  ``Local BRST cohomology in gauge theories,''
  Phys.\ Rept.\  {\bf 338}, 439 (2000)
  [hep-th/0002245].

\bibitem{gravitons}
  N.~Boulanger, T.~Damour, L.~Gualtieri and M.~Henneaux,
  ``Inconsistency of interacting, multigraviton theories,''
  Nucl.\ Phys.\ B {\bf 597}, 127 (2001)  [hep-th/0007220].

\bibitem{N=1SUGRA}
  N.~Boulanger and M.~Esole,
  ``A Note on the uniqueness of D = 4, N=1 supergravity,''
  Class.\ Quant.\ Grav.\  {\bf 19}, 2107 (2002)  [gr-qc/0110072].

\bibitem{HLGR}
  M.~Henneaux, G.~Lucena Gomez and R.~Rahman
  ``Higher-Spin Fermionic Gauge Fields and Their Electromagnetic Coupling,''
  JHEP {\bf 1208}, 093 (2012)
  [arXiv:1206.1048 [hep-th]].



\end{thebibliography}
\end{document}